\documentclass[12pt]{article}
\usepackage{amsfonts}
\usepackage{graphicx}
\usepackage{amsthm}
\usepackage{latexsym,amsmath,color}
\usepackage[round]{natbib}
\usepackage[colorlinks=true,linkcolor=black,urlcolor=black]{hyperref}
\usepackage[english]{babel}
\usepackage{multicol}
\usepackage{epstopdf}
\usepackage{subfigure}
\usepackage{doi}
\usepackage{hyperref}
\usepackage{float}
\usepackage[page]{appendix}
\usepackage{geometry}
\hypersetup{citecolor=black}

\usepackage{array}
\usepackage{tabulary}
\newcolumntype{K}[1]{>{\centering\arraybackslash}p{#1}}

\begin{document}

\title{Testing first-order intensity model in non-homogeneous Poisson point processes with covariates}
\author{M.I. Borrajo, W. Gonz\'alez-Manteiga and M.D. Mart\'inez-Miranda}
\date{}

\newtheorem{theorem}{Theorem}[section]
\newtheorem{conjecture}[theorem]{Conjecture}
\newtheorem{corollary}[theorem]{Corollary}
\newtheorem{lemma}[theorem]{Lemma}
\newtheorem{proposition}[theorem]{Proposition}

\newtheorem*{remark}{Remark}

\newtheorem{definition}{Definition}

\newcommand{\trdsegx}{tr(HD^2\lambda_0(x))}
\newcommand{\trdsegy}{tr(HD^2\lambda_0(y))}
\newcommand{\trdsegxc}{tr^2(HD^2\lambda_0(x))}
\newcommand{\trdsegyc}{tr^2(HD^2\lambda_0(y))}
\newcommand{\KK}{(\boldsymbol{K}\circ \boldsymbol{K})(H^{-1/2}(x-y))}
\newcommand{\KKc}{(\boldsymbol{K}^2\circ \boldsymbol{K})(H^{-1/2}(x-y))}
\newcommand{\KKcc}{(\boldsymbol{K}^2\circ \boldsymbol{K}^2)(H^{-1/2}(x-y))}

\allowdisplaybreaks

\maketitle

\begin{abstract}
Modelling the first-order intensity function is one of the main aims in point process theory, and it has been approached so far from different perspectives. One appealing model describes the intensity as a function of a spatial covariate. In the recent literature, estimation theory and several applications have been developed assuming this model, but without formally checking this assumption.
In this paper we address this problem for a non-homogeneous Poisson point process, by proposing a new test based on an $L^2$-distance. We also prove the asymptotic normality of the statistic and we suggest a bootstrap procedure to accomplish the calibration.  Two applications with real data are presented and a simulation study to better understand the performance of our proposals is accomplished. Finally some possible extensions of the present work to non-Poisson processes and to a multi-dimensional covariate context are detailed.	
\end{abstract}

\section{Introduction}
The understanding of the spatial distribution of point patterns is crucial in many diffe\-rent fields: ecology (\cite{Illian2009} and \cite{Law2009}); epidemiology (\cite{Lawson2013}); seismology (\cite{OgataZhuang2006} and \cite{Schoenberg2011}); forestry (\cite{Stoyan2000}) and geology (\cite{foxall2002nonparametric}). Point processes have also been of interest in methodological statistics, see for example \cite{DaleyVereJones1988} focused on a measure theory approach, and \cite{Diggle2013} and \cite{Cressiebook} that propose a more exploratory perspective.

Initially, from the point of view of statistical inference, the main interest of spatial point process theory was to test the complete spatial randomness (CSR) hypothesis, i.e., to determine if a pattern come from a homogeneous Poisson process, in which case the underlying intensity is constant on the observation region and there is no interaction among points. Distance-based and quadrat  counts are the classical methods on this topic, see \cite{Diggle2013} and \cite{Cressiebook} for more details. Some applications to real situations have also been developed under this assumption, as it can be seen in  \cite{Baddeley1995} and \cite{Dasgupta1998}.

First-order characteristics, among which the intensity function is one of the most important, have been studied from different perspectives. The first-order intensity function is formally defined as:
\begin{equation*}
	\lambda(x)=\lim_{|dx|\to 0}\frac{E[N(dx)]}{|dx|}, 
\end{equation*}
where $N$ is the counting measure, i.e., $N$ is the random variable that counts the number of points lying on a given set, $E$ denotes the expectation and $|\cdot|$ is the Lebesgue measure of the corresponding space (length, area, volume, ...).

The estimation of the first-order intensity function was one of the first problems addressed from methodological statistics. A classical approach consists of estimating the intensity function assuming a parametric model. Parametric fitting can be performed using for instance a likelihood score (see \cite{Waagepetersen2007}, \cite{Moller2003} and \cite{Diggle2013}), or pseudolikelihood procedures, see \cite{Vanlis2000}. However these parametric solutions may generate unreliable estimates which do not represent the underlying true model. Hence the non-parametric approach is a valuable and powerful alternative. \cite{Diggle1985} proposed the first kernel intensity estimator which has been widely used in exploratory analysis. The main disadvantage of this estimator is its lack of consistency, for which \cite{CucalaThesis} proposed an improvement based on the relationship between intensity and density functions, specifically both are non-negative functions and the difference relies only on the fact that the intensity function does not necessarily integrates to one.

Recently, this lack of consistency and some real application requirements induced a new scenario based on the inclusion of covariates in the model. Considering spatially va\-rying covariates has been a big step forward on point process theory and it has been mostly addressed so far from a parametric perspective. See for example \cite{Waagepetersen2007} where they do inference on Neymann-Scott processes depending on a linear combination of spatial covariates; \cite{GuanLoh2007} that estimates the coefficients of the covariates using the Poisson likelihood estimator, previously proposed in \cite{Schoenberg2005}, and also proves the asymptotic normality of those estimates for a general class of mixing processes; and \cite{GuanShen2010} that proposes a method to estimate the coefficients of the covariates improving the Poisson likelihood estimator.

In the non-parametric context the inclusion of covariates has been less prolific. \cite{Guan2008} proposed a kernel intensity estimator, assuming that the intensity function depends on some observed spatially varying covariates through an unknown continuous function. The consistency of his estimator was proved under an increasing domain asymptotic framework, and he also deals with the problem of high-dimensional covariates using sliced inverse regression. Later, \cite{Baddeley2012} postulates that
\begin{equation}\label{eq:modelint}
\lambda(x)=\rho(Z(x)), \: x \in W\subset \mathbb{R}^2,
\end{equation}
where $Z:W\subset \mathbb{R}^2 \rightarrow \mathbb{R}$ is a spatial continuous covariate that is exactly known in every point of the region of interest $W$. The proposal provides with some intensity estimators based on local likelihood, as well as some others based on kernel theory, but without in\-clu\-ding any theoretical developments. \cite{Borrajo_Intensity} detailed the theoretical properties of a kernel intensity estimator in the context of point process with covariates; proposing a specifically designed bootstrap procedure and innovative data-driven bandwidth selectors. 

A little attention has been paid to test the significance of these covariates or the goodness-of-fit of these models. There has been some proposals in the parametric approach for model selection, see \cite{YueLoh2015} for Neymann-Scott processes and \cite{ThurmanGuan2015} for cluster patterns. In the non-parametric field, to the extent of our knowledge, only something similar has been done in a slightly different context in \cite{DiazAvalosMateu2014}, where the authors assume that the intensity depends on a linear combination of several covariates and they test, using the conditional intensity function, if any of the coefficients associates to the covariates may be null.

In this work we try to fulfil the existing gap on checking the goodness-of-fit of model \eqref{eq:modelint} for Poisson point processes by defining a new testing procedure, based on the comparison through $L^2$-distance of the model based on the covariate and the ``classical non-parametric model'', that just takes into account the spatial locations. Note that as we set a Poisson assumption, the point process is completely characterized by its first-order intensity function and the testing procedure is well defined. 

The rest of this paper is organised as follows. Section 2  is devoted to present in detail two motivating examples based on real data, which we use all along the manuscript. In Section 3 a brief introduction to some details on intensity estimation is included and the statistical test is introduced, detailing its asymptotic distribution and a bootstrap procedure used to perform its calibration. The proposed methodology is applied to the real data sets in Section 4 and the performance of the test is analysed in Section 5 through an extensive simulation study, where simulated models have been chosen to reproduce the real data sets in order to make it more realistic. We include some extensions to the present work in Section 6 and we finally draw the overall conclusions in Section 7.


\section{Motivating examples}
In this section we first motivate our proposal using two real data set. The first one is the Murchison data set that has been  used by A. Baddeley in his papers, books and software, see for example \cite{Baddeley2012} and \cite{Baddeley2015}. The second data set has been made up of wildfires in Canada and has been obtained from the Canadian Wildland Fire Information System \texttt{http://cwfis.cfs.nrcan.gc.ca/home}.

\begin{figure}[H]
	\centering
	\includegraphics[scale=0.7]{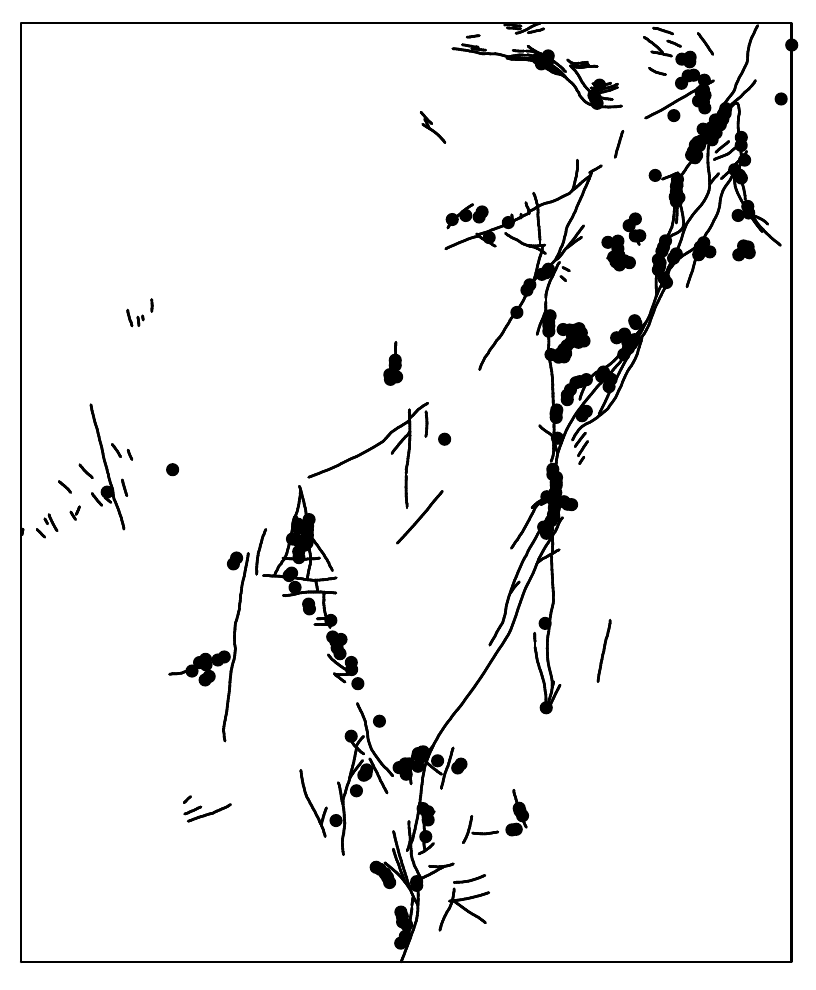}
	\caption{Murchison geological survey data; gold deposits (points) and geological faults (lines).}\label{fig:murchini}
\end{figure}

Murchison geological survey data shown in  Figure \ref{fig:murchini} record the spatial locations of gold deposits (a total number of 255) and the surrounding geological faults. These data came from a $330\times 394$ km region in the Murchison area of Western Australia and they have been obtained by \cite{MurchisonSurvey}. As it has been remarked in \cite{Baddeley2012}, at this scale (1:500000) the gold deposits spatial extension is negligible and they can be considered as points without losing generality; note also that the real gold deposits and faults are three-dimensional while here we use a two-dimensional projection. Moreover, some geological faults may have been missed because they are not recorded by direct observation but in magnetic field surveys or geologically inferred from discontinuities in the rock sequences. 

\begin{figure}[H]
	\centering
	\includegraphics[scale=0.5]{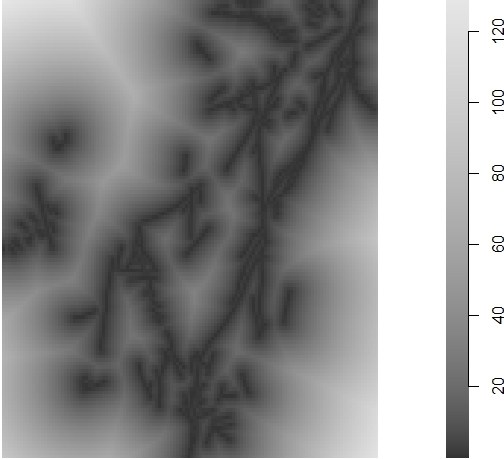}
	\caption{Covariate information for the Murchison data set: distance to the nearest geological fault (in meters).}\label{fig:Murch_DistFaults}
\end{figure}

Once we have the locations of the gold deposits and the faults, the construction of the covariate is simple, we compute the distance from every point in the observation region to the nearest fault (see Figure \ref{fig:Murch_DistFaults}). This covariate is used to model the intensity of the process through the model specification given in \eqref{eq:modelint}.

In \cite{Baddeley2012} the aim on studying these data is to ``specify zones of high prospectivity to be explored for gold'', so they have already assumed that the influence of the fault information is relevant, and it may actually explain the localisation of gold deposits under model \eqref{eq:modelint}. Our goal is to check the adequacy of this model and test the hypothesis that the distance to geological faults is enough to explain the spatial distribution of gold deposits, i.e., the intensity of the process can be explained using the information given by the distance to the nearest geological fault  in the way shown in \eqref{eq:modelint}.

\begin{figure}[H]
	\centering
	\includegraphics[scale=0.6]{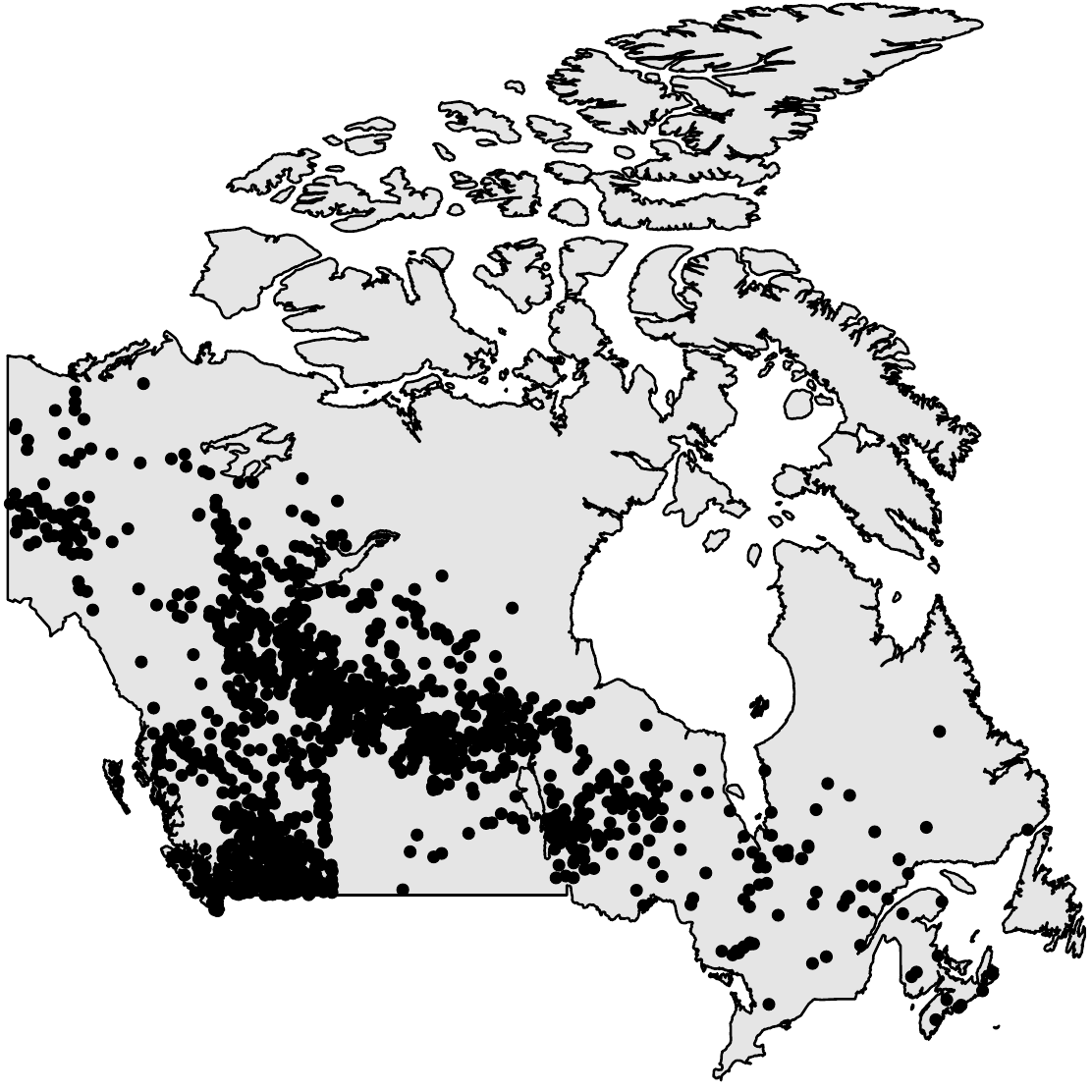}
	\caption{Wildfires in Canada during June 2015.}\label{fig:Canadaini}
\end{figure}

Forest fires are one of the most important natural disturbances since the last Ice Age and they represent a huge social and economic problem. Canada has quite a long tradition on recording information about their wildfires; and also studies from many different perspectives have been carried out: \cite{CanadaFires_Clouds}, \cite{CanadaFires_HighLatitudeCooling}, \cite{CanadaFires_SmokeMediterraneanSea}, \cite{CanadaFires_Meteorological}. It is quite well known that fire activity in Canada mostly relies on meteorological elements such as long periods without rain and high temperatures. In this context we are interested in studying the spatial influence of these meteorological variables on the distribution of wildfires following model \eqref{eq:modelint}. 

\begin{figure}[H]
	\centering 
	\includegraphics[scale=0.7]{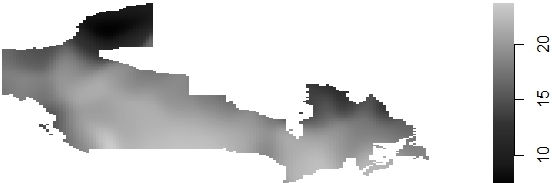}
	\caption{Third quartile of the temperature registered in June 2015 in Canada, after a Gaussian smoothing with $\sigma=2$ (in Celsius degrees).}\label{fig:Canadatemp}
\end{figure}

The wildfire data set and also a complete me\-teo\-ro\-lo\-gi\-cal information from the last decades is available at the Canadian Wildland Fire Information System website (\url{http://cwfis.cfs.nrcan.gc.ca/home}). The fire season in Canada lasts from late April until August, with a peak of activity in June and July, hence we analyse the influence of meteorological covariates on wildfires during June 2015 (a total number of 1841), see Figure \ref{fig:Canadaini}. We will first focus our attention on temperature, see Figure \ref{fig:Canadatemp}, because the above mentioned specific bibliography has determine it as the main necessary factor in the ignition of a wildfire. It is important to note that for inferential purposes we have removed two regions (Northwest Territories and Nunavut, mostly covered by ice layers) from the whole observation window (Canada) because there are no fires registered on those regions and we cannot do any inference with such a lack of information.

Our aim is to test if the distribution of the wildfires in Canada may be explained with the temperature information instead of the spatial locations of the ignition points, i.e., the intensity function of the wildfires may be seen as a function of temperature in the form detailed in \eqref{eq:modelint}. 

To check, on both data sets and in other possible situations, if a selected covariate provides with enough information to determine de model intensity, we present in the next section a testing procedure based on non-parametric techniques for Poisson models, so no second-order interaction between the events is allowed. Remark that in this situation, a point process is completely characterized by its first-order intensity function and the test is then well defined.

\section{The proposed method}\label{sec:test}

\subsection{Previous background}
Our main aim in this work is to propose a goodness-of-fit test for model \eqref{eq:modelint} for Poisson point processes. However, this test will rely on several concepts of non-parametric intensity estimation theory, that we try to briefly introduce and sum up in the present subsection. 

Let $X$ be a Poisson point process defined in a region $W\subset\mathbb{R}^2$, where $W$ is assumed to have finite positive area. The problem may analogously be defined in $\mathbb{R}^n$ but we have reduced it to the spatial context for practical purposes. Let $X_1, \ldots, X_N$ be a realisation of the process where $N$ is the random variable counting the number of events. Let again $Z:W\subset \mathbb{R}^2 \rightarrow \mathbb{R}$ be the spatial continuous covariate that is exactly known in every point of the region of interest $W$. In practice this covariate will commonly be known in an enough amount of points spread on the region, so the values for the rest of the points can be interpolated and it can be assumed that these values are indeed the real ones.

\cite{Diggle1985} proposed the first kernel intensity estimator for one-dimensional point processes, which has been easily extended to the plane:
\begin{equation*}
\hat{\lambda}^D_H(x)=\frac{\sum_{i=1}^N\mathbf{K}_H\left(x-X_i\right)}{p_H(x)}, \quad x \in \mathbb{R}^2,
\end{equation*}
where $H\equiv H(m)$ is a bandwidth matrix, $\mathbf{K}$ denotes a bivariate kernel function, $\mathbf{K}_H(x)=|H|^{-1/2}\mathbf{K}\left(H^{-1/2}x\right)$ and $p_H=\int_W{|H|^{-1/2}}\mathbf{K}(H^{-1/2}(x-y))dy$ is an edge correction term.

This estimator has been widely used during decades for exploratory a\-na\-ly\-sis, but the inference performed with it has been limited due to its lack of consistency. One approach to overcome this problem was proposed in \cite{CucalaThesis}, where the concept of ``density of events locations'' appears $\lambda_0(x)=\lambda(x)/m$, with $m=\int_W{\lambda(x)dx}$ the expected number of events lying on W. He proposes a kernel estimator and he proves its consistency under an infill structure asymptotic framework. \cite{Isa2015} extended these ideas to the two-dimensional situation using bandwidth matrices:
\begin{equation}\label{eq:isaest}
\hat{\lambda}_{0,H}(x)=\frac{\hat{\lambda}^D_H(x)}{N}1_{\{N\neq 0\}}= \frac{1}{p_H(x)N}|H|^{-1/2}\sum_{i=1}^N \mathbf{K}\left(H^{-1/2}(x-X_i)\right)1_{\{N\neq0\}}, x\in\mathbb{R}
\end{equation}
with $|H|$ denoting the determinant and $H^{-1/2}$ a power of matrix $H$.

Another approach to address the problem of the lack of consistency was including covariates in the model. Following this idea and model \eqref{eq:modelint}, in \cite{Borrajo_Intensity} we found a kernel intensity estimator using covariate information, which combines the idea of the ``density of events location'' and the use of covariates in the model postulated by \cite{Baddeley2012}.

The grounds of this proposal lies on a result stating that, if $X$ is a point process in $W\subset \mathbb{R}^2$ fulfilling equation \eqref{eq:modelint} and $Z:W\subset \mathbb{R}^2 \rightarrow \mathbb{R}$  the spatial continuous covariate, then the transformed point process $Z(X)$ is a point process in $\mathbb{R}$ with intensity $\rho g^\star$. This function $g^\star(\cdot)=|W|g(\cdot)$, where $g$ is the derivative of the spatial cumulative distribution function of $Z$, i.e., $G^{'}(z)=g(z)$ with $G(z)=\frac{1}{|W|}\int_W{1_{\{Z(u)\leq z\}}du}$. Hence, the density of events locations associated to the transformed point process is $f(\cdot)=\frac{\rho(\cdot)g^\star(\cdot)}{m}$ with $m=\int_{\mathbb{R}}\rho(z)g^\star(z)dz=\int_W{\lambda(x)dx}$.

Then, \cite{Borrajo_Intensity} defined:
\begin{equation}\label{eq:dens_est}
\hat{f}_{h}(z)=g^\star(z) \frac{1}{N}\sum_{i=1}^N\frac{1}{g^\star(Z_i)}K_h\left(z-Z_i\right)1_{\{N\neq 0\}},
\end{equation}
where $K$ is a univariate kernel function and $K_h(\cdot)=\frac{1}{h}K\left(\frac{\cdot}{h}\right)$. For those familiar with weighted or biased data, this proposal is structurally similar to the kernel density estimator for length biased data proposed by \cite{Jones1991}, and later developed in \cite{Borrajo_LengthBiased} with a complete asymptotical framework, specific bandwidth selection methods and a bootstrap procedure. 

This is an estimate of $f$, and once we have it, we can go back to the intensity function just by plug it in: obtaining an estimator of $\rho$ from the definition of the density of event location, and then using equation \eqref{eq:modelint} to get an estimate of $\lambda$.

\subsection{The test}
Formally, we want to test a null hypothesis $H_0: \lambda(x)=\rho(Z(x)), \, x \in W$ versus a general alternative in which the intensity function is not explained completely through the covariate, for Poisson processes. Recall that we do not require any condition on $\rho$ more than being a one-dimensional real-valued function. Note also that, as we are under a Poisson assumption, the process under the null hypothesis is well-defined and absolutely characterised, see Section 6 for some ideas on a more general version of our proposal in terms of the nature of the process.

The idea is to define a test statistic based on a $L^2$-distance between the classical kernel intensity estimator using only location information and the appealing one using covariate information. To avoid the problem of the lack of consistency we are using the density of event location instead of the corresponding intensities. Hence, the null hypothesis can be equivalently rewritten as $H_0: \lambda_0(x) = \rho(Z(x))/m$, with recall $\lambda_0(x)=\lambda(x)/m$, and $m=\int_W\lambda(x)dx$ denoting the expected number of events in $W$.

Let define the $L^2$-distance between the two theoretical functions of interest, i.e, the relative density without assuming any particular characteristic and the relative density under the covariate influence as stated in \eqref{eq:modelint}:
\[D=\int_W{\left(\lambda(x)/m - \rho(Z(x))/m\right)^2},\]
where $m=\int_W{\lambda(x)dx}=\int_W{\rho(Z(x))dx}$. See \cite{Borrajo_Intensity} for the details on how the transformed process through the covariate preserves the expected number of events lying in $W$, i.e., on how the Poisson condition is inherited.

The procedure to construct the statistic is: we first estimate the relative density with the two-dimensional kernel estimator \eqref{eq:isaest}, just using the event locations information and then we estimate it using \eqref{eq:dens_est}, based only on covariate information. Then we apply the $L^2$-distance to obtain a statistic that measures the discrepancy between the null and the alternative hypotheses. When the distance is big, we reject the null hypothesis, because we have significant evidence in favour of the alternative, i.e., the intensity of the Poisson process cannot be explained by the covariate information in the previously established form; and when it is small, the intensity of the Poisson process may be explained using only the covariate information.

Hence, the test statistic is formally defined as follows:
\begin{equation} \label{eq:Tstat}
T=\int_W{\left(\hat{\lambda}_{0,H}(x)-\hat{\rho}_{0,b}(Z(x))\right)^2dx},
\end{equation}
where $\hat{\rho}_{0,b}(Z(x))=\frac{\hat{\rho}_b(z)}{N}1_{\{N\neq 0\}}$ with $\hat{\rho}_b(z)=\hat{f}_b(z)m/g^\star(z)$, with $b\equiv b(m)$ a real bandwidth parameter, $L$ a univariate kernel function and $L_b(z)=\frac{1}{b}L\left(\frac{z}{b}\right)$, see \cite{Borrajo_Intensity} for more the details on these functions.

\subsection{Asymptotic properties and calibration}
In this section we derive the asymptotic distribution of the statistic \eqref{eq:Tstat} under a suitable asymptotic framework. We find our inspiration on the related context of kernel density estimation where the common asymptotics assume that the deterministic sample size tends to infinity. In point process we may find at least two possibilities. One is the increasing domain, see \cite{Guan2008}, where the expected number of events tends to infinity with the increasing size of the observation region. Remark that with this idea, ``new points'', i.e., extra information, is only given in the boundary of the region and the estimated intensity at each point depends on an expected number of events tending to zero. The second option, initially proposed by \cite{DiggleMarron1988}, is the infill structure asymptotic framework, which overcomes the previously detailed problem stating that the expected number of events tends to infinity for a fixed bounded observation domain. In this work we have used this second option following \cite{Cowling1996}, \cite{CucalaThesis}, \cite{Isa2015} and \cite{Borrajo_Intensity} among others.

At this point we need to introduce some regularity conditions. Recall that the estimator $\hat{\lambda}_{0,H}$ proposed in \cite{Isa2015} uses a bandwidth selector defined in that same work, and its consistency has been proved under an infill structure asymptotic framework, where the authors assume that $W=\mathbb{R}^2$ in order to avoid bias towards undersmoothing, see \cite{Isa2015} and \cite{DiggleMarron1988} for details on this. As these consistency results are required we keep the condition of $W=\mathbb{R}^2$ only for the theoretical developments. 

\begin{itemize}
	\item[(A.1)]{$\int_{\mathbb{R}}L(z)dz=1$; $\int_{\mathbb{R}}zL(z)dz=0$ and $\mu_2(L):=\int_{\mathbb{R}}z^2L(z)dz<\infty$.}
	\item[(A.2)]{$\lim_{m\to\infty}b=0$ and $\lim_{m\to\infty}\frac{A(m)}{b}=0$, where $A(m):=\mathbb{E}\left[\frac{1}{N}1_{\{N\neq 0\}}\right]$.}
	\item[(A.3)] The bandwidth matrix $H$ is symmetric and positive-definite, such that all entries of $H$ tends to zero and $m^{-1}|H|^{-1/2} \to 0$ as $m$ increases.
	\item[(A.4)] $\boldsymbol{K}$ is a continuous, symmetric, square integrable bivariate density function such that $\int_{\mathbb{R}^2}{uu^T\boldsymbol{K}(u)du}=\boldsymbol{\mu_2(K)}I_2$ with $\boldsymbol{\mu_2(K)} <  \infty$ and $I_2$ denoting the two-dimensional identity matrix.
	\item[(A.5)]{$Z(x)$ is a continuity point of $\rho$ for all $x \in W$.}
\end{itemize}

The following theorem provides an analogous result to the one in \cite{Hall1984}, i.e., a central limit theorem for the integrated squared error of multivariate kernel density estimators, for the statistic $T$ in  the context of spatial point process with covariates. The proof of this result is detailed in the Appendix of the present work.

\begin{theorem}\label{th:normality}
Under conditions (A.1)  to (A.5) and assuming the null hypothesis, $H_0: \lambda(x)/m = \rho(Z(x))/m \quad \forall x \in W$, it holds that
\[\frac{T - \mu_T}{\sigma_T} \longrightarrow N(0,1),\]
where
\[\mu_T= A(m)|H|^{-1/2}R(\boldsymbol{K}) +\frac{1}{2}\boldsymbol{\mu_2(K)}\int\lambda_0(x)\trdsegx dx+\frac{1}{4}\boldsymbol{\mu_2^2(K)}\int \trdsegxc dx,\]
\[\sigma_T^2= A(m)|H|^{-1/2}\int\int\lambda_0^2(x)\lambda_0(y)\KK dxdy+ 2 A(m)|H|^{-1/2}R(\lambda_0)R(\boldsymbol{K}),\]
with $\circ$ denoting the convolution between two functions, $tr(\cdot)$ the trace of a matrix, $D^2$ the Hessian matrix  and $R(f)=\int f^2(x)dx$ for any function $f$.
\end{theorem} 

In practice, the asymptotic distribution given in Theorem \ref{th:normality} could be approximated estimating $m$ by the sample size $n$, and $A(m)$ by $1/n$, as it has been extensively justified in \cite{CucalaThesis}. However, this asymptotic distribution may not be the best way to calibrate our test since the convergence rate is too slow and therefore it is not suitable for small patterns. There exists several examples in different context with a similar problem, see \cite{Duong2013} for the two sample problem in density estimation, \cite{EduIngrid} for directional data and \cite{IsaSpatStat} for spatial point process without covariates among others. Our proposal is to use bootstrap to calibrate the test.

We have chosen a smooth bootstrap procedure inspired in \cite{Cao1993} and \cite{Cowling1996} to resample under the null hypothesis using the Poisson assumption as we have previously remarked. In  Section 6 we provide ideas for extensions to a more general situation.  
Under the null hypothesis the intensity can be written as a function of the covariate, $\lambda(x)=\rho(Z(x))$. So let consider our pilot estimate $\hat{\lambda}(x)=\hat{\rho}_t(Z(x))$, with pilot bandwidth  $t$. With this pilot estimate and conditional on the observed pattern $X_1,\ldots,X_N$, the bootstrap algorithm for the calibration of the test consists of the following steps.
\begin{enumerate}
 \item Draw the number of events, $n^\ast$, from a random variable $N^\ast$ having a Poisson distribution with intensity $\int_W{\hat{\rho}_t(Z(x))dx}$.

\item Draw the bootstrap sample, $X_1^\ast, \ldots,X^\ast_{n^\ast}$, by sampling randomly $n^\ast$ times from the distribution with density proportional to $\hat{\lambda}(x)=\hat{\rho}_t(Z(x))$. 

\item Compute the test statistic under the bootstrap distribution, $T^\ast$, using the expression in \eqref{eq:Tstat} applied to the bootstrap sample. 

\item Repeat steps 1 to 3 $B$ times to obtain the empirical quantile of the statistic. 
\end{enumerate}
The test would conclude by comparing the value of the statistic using the original data with the obtained empirical quantile to decide whether reject or not the null hypothesis.

%
Apart from the pilot bandwidth $t$, the testing procedure involves the choice of several bandwidth parameters. Optimal bandwidth selection for tests based on kernel smoothing is still an open problem. Nevertheless in practical situations is often suitable to use good automatic bandwidth selectors designed for estimation, even though they might not be theoretically optimal for testing purposes. In point processes this has been considered for example in the former paper by \cite{Isa2015}; in the density field for testing different characteristics, see for example \cite{Silverman1981} and \cite{MammenMarronFisher1992} for multimodality tests, \cite{Hyndman2002} for symmetry tests and in the regression context see \cite{Hart1992} for goodness-of-fit. To derive our test we need a bandwidth matrix $H$ for the intensity estimator of \cite{Isa2015}, as well as a scalar bandwidth for the covariate-based intensity estimator of \cite{Borrajo_Intensity}. Following the precedent literature we use the automatic data-driven bandwidth selectors proposed for the estimators in \cite{Isa2015} and  \cite{Borrajo_Intensity}, respectively. In the numerical analyses showed later we will see that these choices seem to be appropriate, while for the pilot bandwidth $t$ we have evaluated the test in a suitable range of values to validate our results. 

\section{Data analyses}\label{sec:dataanal}
In this section we apply to real data the proposed test to check the intensity dependence on the covariate in the form of \eqref{eq:modelint}, assuming a spatial Poisson point process. As mentioned in Section 2, our real data examples come from two sources, a geological survey in the Murchison area of Western Australia and wildfire tracking by the national Wildland Fire Information System in Canada. The test is applied to decide if, in the first case, the distance to geological faults explains by itself (without the locations) the spatial distribution of gold deposits, and similarly in the second one for temperature and wildfires in Canada during June 2015. 

Taking into account a previous remark, our test statistic relies on several bandwidth parameters that needs to be selected. We propose to use automatic data-driven bandwidth selectors for the two involved in the computation of the test statistic in \eqref{eq:Tstat}, which are a scalar bandwidth $b$, and a matrix of bandwidths $H$. These selectors are respectively the methods of \cite{Borrajo_Intensity} and  \cite{Isa2015}. With these choices it only remains one more scalar bandwidth to be determined, this is the pilot bandwidth $t$ used for the bootstrap calibration of the test. We have decided to use different values in a suitable range to thoughtfully validate our results; for each data set the range has been chosen around the automatic choice of the method proposed in \cite{Borrajo_Intensity}, taking into account the scale of the data not to oversmooth nor undersmooth.\\

\noindent \textit{Murchison gold deposits}\\
Recall that here we have the gold deposit locations in an area of 330 km $\times$ 394 km and the distance to the nearest fault as covariate (represented in Figure \ref{fig:Murch_DistFaults}). Taking into account the bandwidth value obtained with the procedure in \cite{Borrajo_Intensity}, we conclude that an interval around 0.52, which is the value of the automatic bandwidth selector, is an appropriate range for the pilot bandwidth parameter $t$ in the bootstrap calibration. Using smaller bandwidths would lead to undersmoothed estimates, with lots of wiggles, while using larger values we are on the opposite end with inappropriate  oversmoothed intensities.

\renewcommand{\arraystretch}{1.4}
\begin{table}[H] 
	\centering
	\begin{tabular}{p{1.8cm}|ccccccc|} \cline{2-8}
		& $t=0.3$ & $t=0.4$ & $t=0.5$ & $t=0.55$ & $t=0.6$ & $t=0.65$ & $t=0.7$ \\ \hline
		\multicolumn{1}{|l|}{p-value}& 0.938 & 0.894 & 0.804 & 0.766 & 0.702 & 0.646 & 0.602 \\ \hline 
	\end{tabular}
	\caption{P-values of the test for different pilot bandwidths in the bootstrap algorithm for the Murchison data set.}\label{tab:Murchreal}
\end{table}

Table \ref{tab:Murchreal} shows the p-values obtained for different pilot bandwidth values in the above detailed range. We have used $B=500$ replications for the bootstrap algorithm and from the results in this table we cannot reject the  null hypothesis for this data set not having enough evidence against. All the bandwidth in the considered range provide high p-values. Hence, once it is assumed that the process fulfils a Poisson condition, there is statistical evidence supporting that the geological faults are enough to explain the location of the gold deposits in the form shown in model \eqref{eq:modelint}. In this particular data set, Poisson assumption seems to be appropriate since the correlation in the data may be negligible, i.e., having a gold deposit at a certain location does not favour or prevent from having another gold deposit nearby.\\

\noindent \textit{Wildfires in Canada}\\
Now, we want to test if the temperature plays a fundamental role in the intensity of the wildfires in Canada during June 2015 in the way we have previously detailed. We use again  the same automatic bandwidth choices as above to compute the value of the test statistic, and different bandwidth values in a suitable range for the pilot bandwidth are used in the bootstrap calibration. For these data it seems that a suitable range would be  around 0.35 which is the value of the automatic selector. The results were the same for all the bandwidths in the range, rejecting the null hypothesis based on p-values which were always close to zero. Even similar conclusions were derived  considering  bandwidths in a wider range outside the original one. Hence, we can conclude that, either there is interaction among the events so the Poisson assumption is not fulfilled causing a high influence on the result of the test, either even the temperature is likely to have an effect on the distribution of the wildfires in Canada and it is not enough to explain the first-order intensity by itself in the way shown in \eqref{eq:modelint}. Actually we think that both ideas are met: on one hand wildfires seem to have interaction among them (a wildfire may even be the cause of the origin of another) and on the other hand the idea of being able to explain wildfires occurrences only with temperature information may be too ambitious. This suggests that maybe another meteorological covariates or some indexes (gathering several variables) should be used to analyse this process. This data set is really suitable to apply some of the extensions detailed later in Section 6.

\section{Simulated illustrative examples}\label{sec:simus}
This section is devoted to analyse the performance of our proposal through Monte Carlo simulations. The models we use are based on the real data sets previously presented in Section 2 and analysed in Section \ref{sec:dataanal}. 

\begin{figure}[H] 
	\centering
	\includegraphics[scale=0.6]{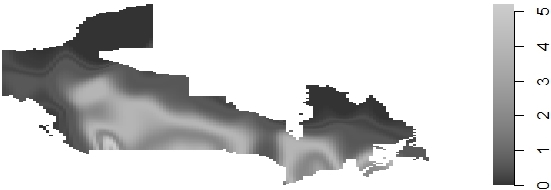}
	\caption{Theoretical intensity function for the first model analysed in the simulation study, that has been obtained applying a kernel intensity estimator to the Canada wildfire data set.}\label{fig:intmodCan}
\end{figure}

The first model is based on the Canada data set, so, our theoretical model is the intensity obtained after applying the kernel intensity estimator proposed in \cite{Borrajo_Intensity} to the wildfire data set during June 2015 with the temperature, see Figure \ref{fig:Canadatemp}, as covariate. This intensity function is represented in Figure \ref{fig:intmodCan}.

\begin{figure}[H] 
	\centering
	\includegraphics[scale=0.5]{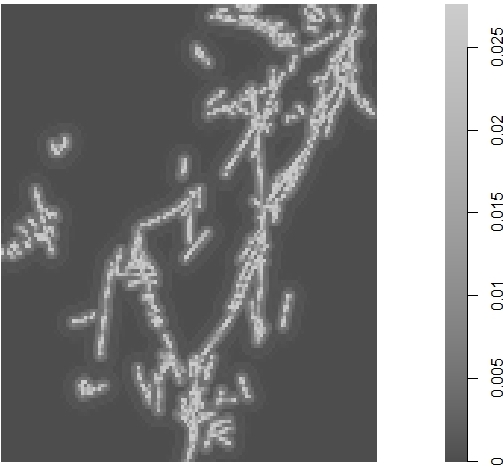}
	\caption{Theoretical intensity function for the second model analysed in the simulation study, that has been obtained applying a kernel intensity estimator to the Murchison data set.}\label{fig:Murch_IntensityModel}
\end{figure}

The second model has been constructed in a similar way but using the Murchison data set. The intensity is represented in Figure \ref{fig:Murch_IntensityModel} and the covariate used is the distance to the nearest geological fault (see Figure \ref{fig:Murch_DistFaults}).

These two models adding the Poisson assumption, lie under the null hypothesis, indeed their intensity function depends on a covariate through a univariate function, in both cases the one given by the expression of the non-parametric kernel intensity estimator of \cite{Borrajo_Intensity}. Hence, it is enough to generate the samples from a non-homogeneous Poisson point process with this intensity function.

To evaluate the power of the test we define multiplicative models, based on the initial ones, that depend on a parameter regulating the discrepancy from the null hypothesis
\[\lambda(x)=\lambda_{\mbox{ini}}(x)r(x),\]
where $\lambda_{\mbox{ini}}$ denote the intensity of the initial model under the null and $r$ is the function to perturb it depending on a parameter. This function corresponds with a diagonal band that crosses the observation region nullifying the extension out of it, with a smooth change. The band is wider or thinner depending on a one-dimensional parameter: when its wide enough it covers the whole observation region, so we approach the null hypothesis, and as it becomes thinner the model gets away from it.

\begin{figure}[H] 
	\centering
	\includegraphics[scale=0.35]{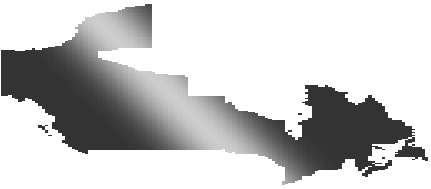}\hspace*{0.4cm}\includegraphics[scale=0.35]{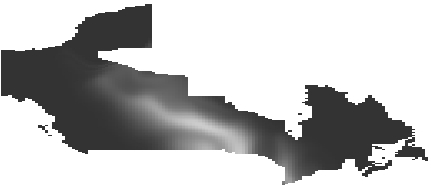}
	\vspace*{0.2cm}
	
	\includegraphics[scale=0.35]{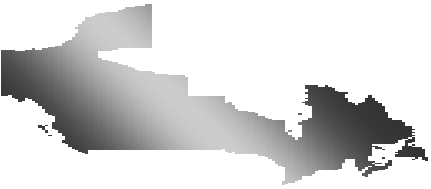}\hspace*{0.4cm}\includegraphics[scale=0.35]{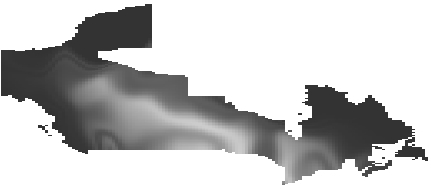}
	\vspace*{0.2cm}
	
	\includegraphics[scale=0.35]{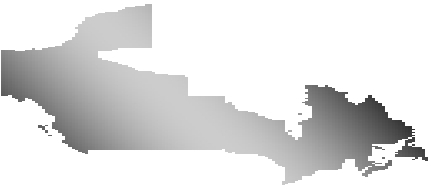}\hspace*{0.4cm}\includegraphics[scale=0.35]{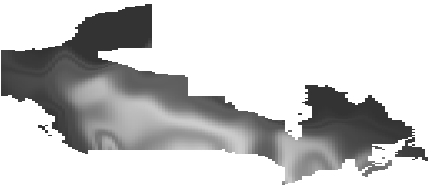}
	\vspace*{0.2cm}
	
	\includegraphics[scale=0.35]{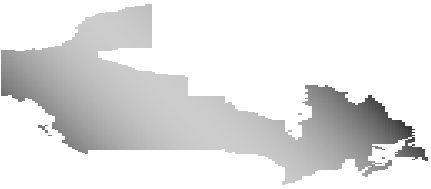}\hspace*{0.4cm}\includegraphics[scale=0.35]{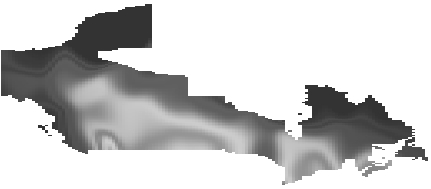}
	\caption{Representation of the $r_C$ functions (first column) and the resulting intensity (second column) for the four values of the parameter $d_C=6,12,20,30$ for the Canada model.}\label{fig:Canadalternative}
\end{figure}

We have to define two different functions, because the observation regions of each model are different. However we preserve the same idea, so, in both cases the diagonal band is based on a univariate Normal density, $\phi$, depending on a parameter, $d_C$ in the Canada data set and $d_M$ in the Murchison data set. For the Canada model, $r_C(u,v)=\phi(u,15-v-v_{0C},d_C)$, where $v_{0C}=60.40$ is the middle point of the y-axis in the observation region, and $d_C$ is the parameter which takes the values $6,12,20$ and $30$. In Figure \ref{fig:Canadalternative} we represent, for each value of the parameter $d_C$, both the $r_C$ function (first column) and the final intensity function after multiplying the initial one by $r_C$ (second column).

%

\begin{figure}[H] 
	\centering
	\includegraphics[scale=0.23]{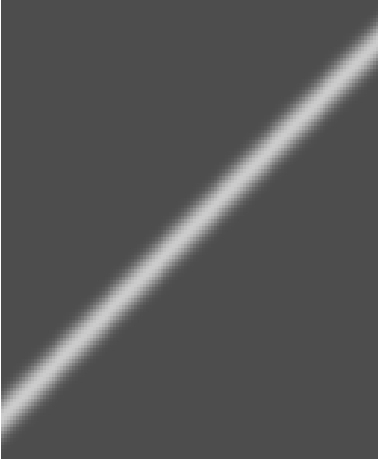}\hspace*{0.3cm}
	\includegraphics[scale=0.23]{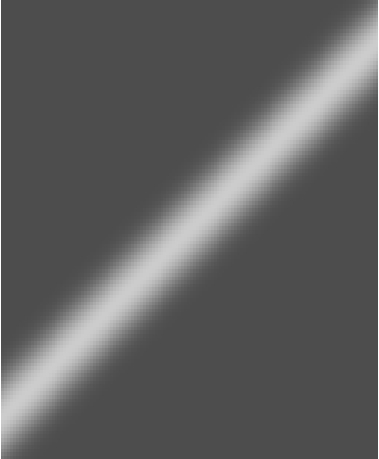}\hspace*{0.3cm}
	\includegraphics[scale=0.23]{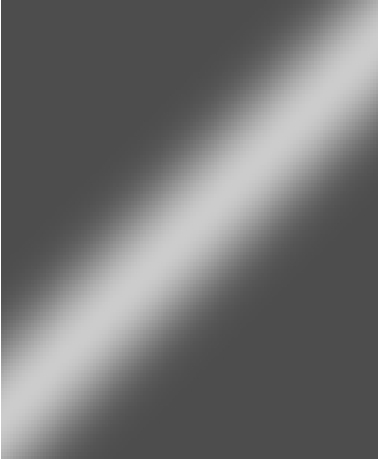}\hspace*{0.3cm}
	\includegraphics[scale=0.23]{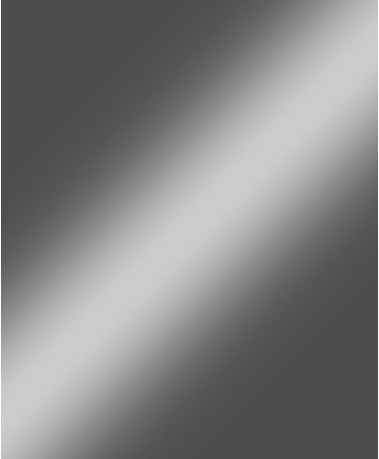}\vspace*{0.5cm}
	
	\includegraphics[scale=0.23]{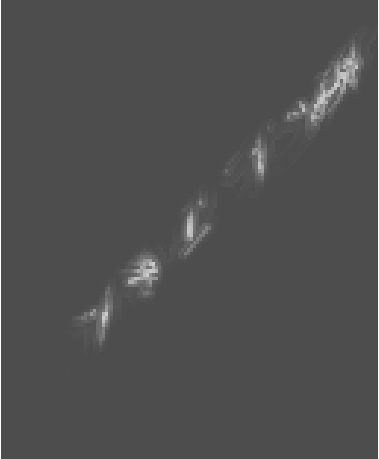}\hspace*{0.3cm}
	\includegraphics[scale=0.23]{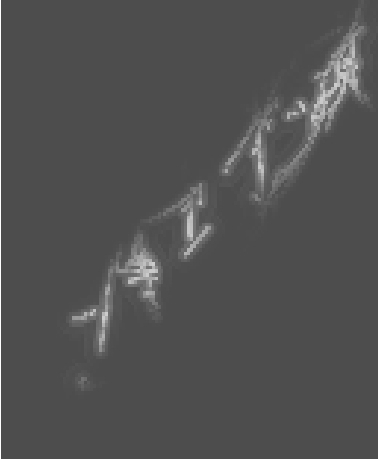}\hspace*{0.3cm}
	\includegraphics[scale=0.23]{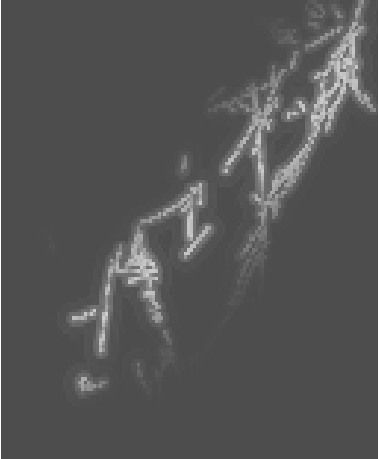}\hspace*{0.3cm}
	\includegraphics[scale=0.23]{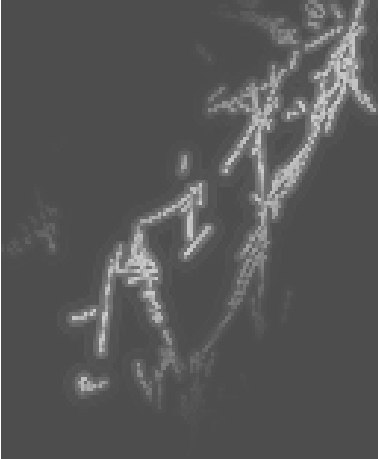}
	\caption{Representation of the $r_M$ functions (first row) and the resulting intensity (second row) for the four values of the parameter $d_M=10,20,40,60$ for the Murchison model.}\label{fig:Murchalternative}
\end{figure}

For the Murchison model, the function is defined as $r_M(u,v)=\phi(u-u_{0M},u-v_{0M},d_M)$, where $u_{0M}=517.69$ and $v_{0M}=6900.61$ are the middle point of the x and y-axis respectively, and $d_M$ is the parameter that, in this case, takes values $10,20,40$ and $60$. We have decided to rotate the diagonal band due to the distribution of the data and the covariate information, but it could have been done in the other way. Take also into account that due to this rotation, even a thin band collect a lot of information from the covariate which may cause smaller rejection proportions. Indeed, we have performed the same simulations with a rotated band and we obtained higher rejection proportions. In Figure \ref{fig:Murchalternative} we represent the four $r_M$ functions (first row) as well as the associated intensities (second row).

Remark that we have not included the scales in Figure \ref{fig:Canadalternative} and \ref{fig:Murchalternative} because the values of the intensity depend on the expected sample size so they will change in each of the situations considered in the study. In the tables below we denote by $d_C=\infty$ and $d_M=\infty$ the initial model without any band restriction on them, i.e., the situation fulfilling the null hypothesis.

\renewcommand{\arraystretch}{1.4}
\begin{table}[H] 
	\centering
	\begin{tabular}{l|ccccc}
		& $d_C=6$ & $d_C=12$ & $d_C=20$ & $d_C=30$ & $d_C=\infty$\\ \hline
		$m=50$ & 1 & 0.6852 & 0.1454 & 0.0722 & 0.0480 \\
		$m=100$ & 1 & 0.9308 & 0.2232 & 0.0826 & 0.0502\\
		$m=200$ & 1 & 0.9986 & 0.4076 & 0.1060 & 0.0516\\
		$m=500$ & 1 & 1 & 0.8266 & 0.1744 & 0.0520\\
		
	\end{tabular}
	\caption{Rejection proportions for the Canada model, with different values of the parameter controlling the discrepancy from the null hypothesis, $d_C$, and four expected sample sizes, $m$.} \label{tab:canadasim}
\end{table}

\renewcommand{\arraystretch}{1.4}
\begin{table}[H] 
	\centering
	\begin{tabular}{l|ccccc}
		& $d_M=10$ & $d_M=20$ & $d_M=40$ & $d_M=60$ & $d_M=\infty$\\ \hline
		$m=50$ & 0.9584 & 0.3858 & 0.1048 & 0.0636 & 0.049\\
		$m=100$ & 0.9926 & 0.4774 & 0.1049 & 0.0680 & 0.0496\\
		$m=200$ & 0.9998 & 0.6404 & 0.1136 & 0.0610 & 0.0504\\
		$m=500$ & 1 & 0.9376 & 0.1727 & 0.0633 & 0.0492\\	
	\end{tabular}
	\caption{Rejection proportions for the Murchison model, with different values of the parameter controlling the discrepancy from the null hypothesis, $d_M$, and four expected sample sizes, $m$.}\label{tab:murchsim}
\end{table}

In Table \ref{tab:canadasim} and Table \ref{tab:murchsim} we show the rejection proportions for different situations that go, from the null hypothesis ($d_{\bullet}=\infty$) to the furthest away situation from it ($d_C=6$ and $d_M=10$, respectively). These rejection proportions have been computed with 5000 Monte Carlo replications for each scenario using 200 bootstrap resamples for the calibration of the statistic. Note that we have carried out this simulation study with different number of bootstrap resamples, realising that from 200 on the results become stable. Regarding the bandwidths values, we follow the same idea we have already used for the real data, so, we use data-driven selectors, see \cite{Isa2015} and \cite{Borrajo_Intensity}, for the computation of the statistic and a pilot bandwidth in a suitable range around the value given by the automatic selector taking the variability into account.

We can see that the power of the test seem to be better for the first model, where even for $d_C=20$, which is a situation near to the null, the power values are high for medium and large expected sample sizes. In the second model, the values do not reach those levels. This is due to the reason we have already pointed out that, if we keep in mind the spatial distribution of the gold deposits, even for thin bands we are gathering a lot of information from the covariate, and hence we are not really as far from the null hypothesis as we may think.

\begin{figure}[H]
	\hspace*{-0.6cm}	\includegraphics[scale=0.45]{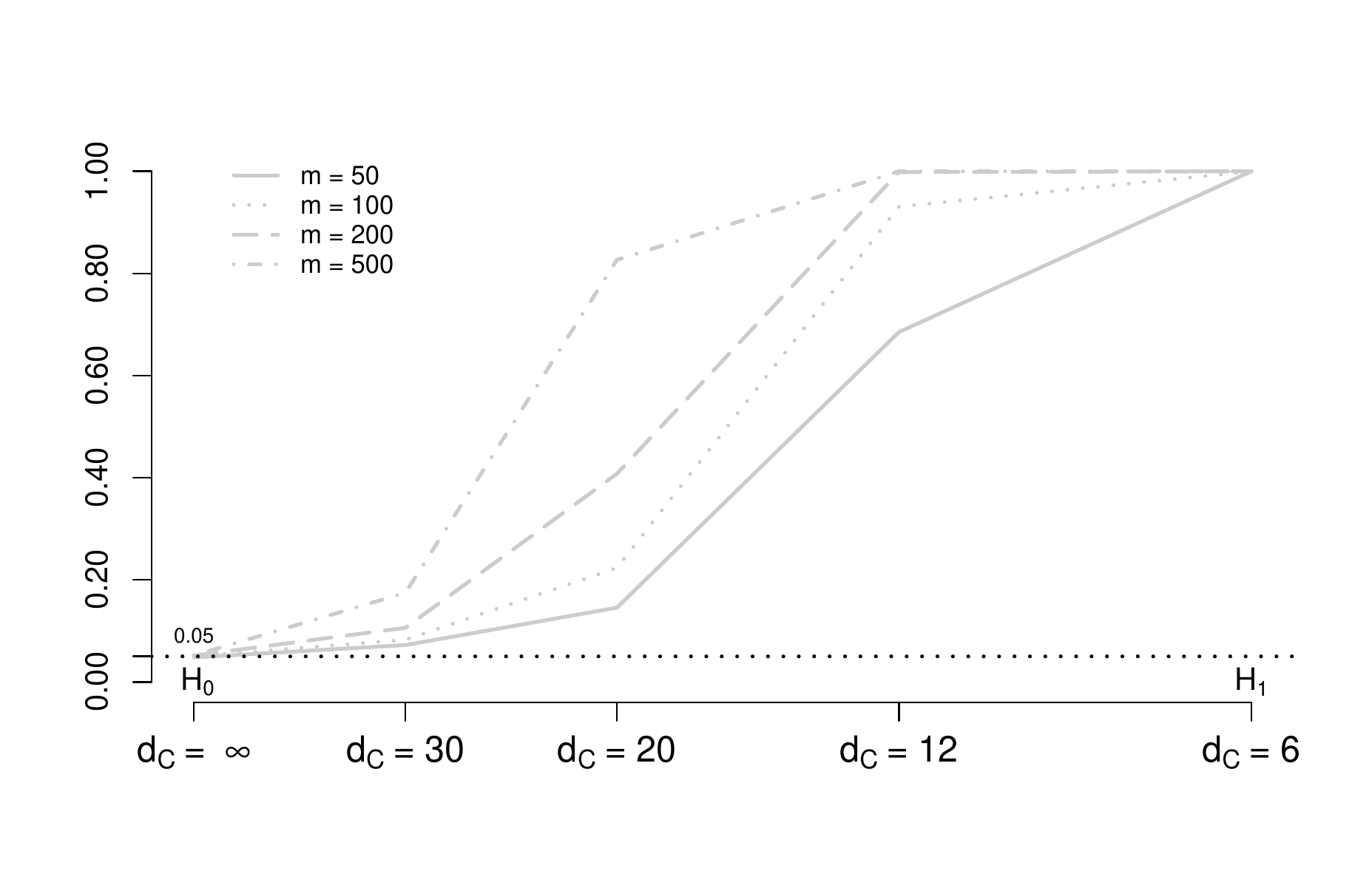}
	\hspace*{-0.4cm}\includegraphics[scale=0.45]{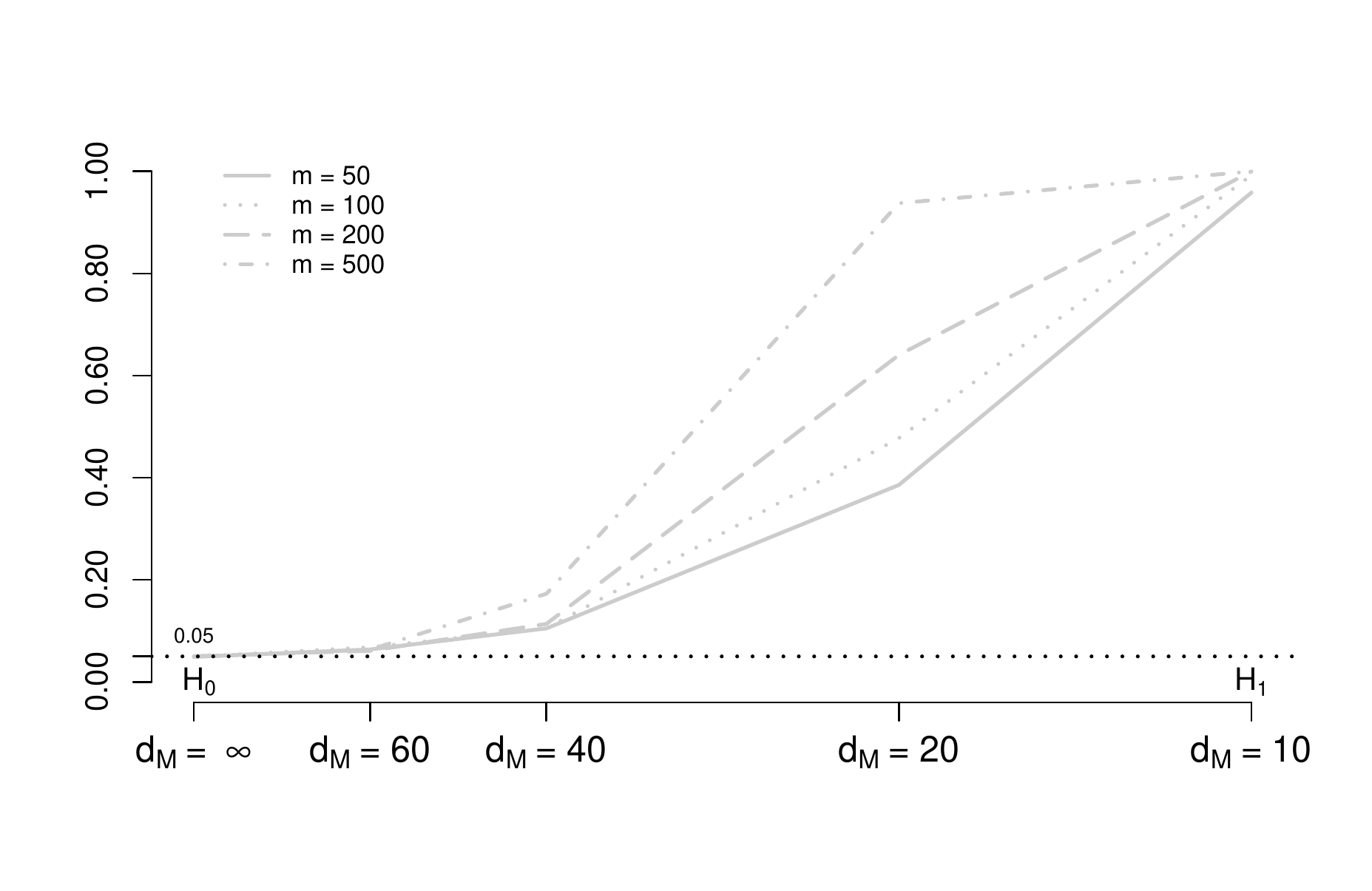}
	\vspace*{-1cm}\caption{Representation of the rejection proportions for the different simulated scenarios of the Canada model (left) and Murchison model (right).}\label{fig:both}
\end{figure}

In order to have a quick visualisation, we have represent those rejection proportion on a graph depending on the distance to the null hypothesis and the sample size, see Figure \ref{fig:both}. 


For both models, Figure \ref{fig:both} shows that in all the scenarios under the null hypothesis the proportion of rejections is around 0.05 (the significance level). This proportion increases as the scenarios get away from the null and, as expected, it increases faster with the growth of the sample size. We can also detect here the fact that for the Murchison model the rejection proportions are generally lower than for the Canada model, due to the relationship between the orientation of the band and the distribution of the data that we have previously remarked.

Note that we have not included any other method to compare with in this study because, to the extent of our knowledge, there is no existing procedure addressing the same problem, so no competitors are available.

\section{Extensions}
Our proposal, as described in the previous sections, is limited to a model with only a one-dimensional covariate; and the goodness-of-fit test of this model relies on the Poisson assumption. These might be important restrictions of our proposal in many real applications, for which we provide in this section with basis ideas for possible extensions. We first address the inclusion of multidimensional covariates in the model, with the corresponding increasing complexity in the related theoretical framework, and then we detail the ideas to extend our proposal to non-Poisson processes. 

\subsection{Multi-dimensional covariates}
We have established $Z:W\subset \mathbb{R}^2 \rightarrow \mathbb{R}$ a continuous covariate. Let now assume that we have $\mathbf{Z}=(Z_1,\ldots,Z_p): W\subset \mathbb{R}^2 \rightarrow \mathbb{R}^p$ a $p$-dimensional continuous covariate where each $Z_i$ fulfils the same smoothness conditions detailed in Section 3 for $Z$.

A first approach to include this $p$-dimensional covariate is try to define a one-dimensional covariate gathering similar information, using for example principal component analysis, defining a combination/index, and applying then the results shown in the previous sections. However this may be not enough in some real complex problems, so we propose to extend the definition of our test to dimension $p$. To do so we need to take into account the formalisation of the transformed process through a p-dimensional covariate detailed in \cite{Borrajo_Intensity}, which guarantee the extension of the transformed process and the corresponding intensity estimator for the multivariate case. Once the transformed multi-dimensional process has been characterised, we define the statistic based on a $L^2$-distance as:
\[\mathbf{T}=\int_W{\left(\hat{\lambda}_{0,H}(x)-\hat{\rho}_{0,M}(Z(x))\right)^2dx},\]
where $\hat{\rho}_{0,M}(Z(x))=\frac{\hat{\rho}_M(x)}{N}1_{\{N\neq 0\}}$ with $\hat{\rho}_M(x)=\sum_{i=1}^N\frac{1}{g^\star(Z(X_i))}\mathbf{L}_M\left(Z(x)-Z(X_i)\right)$, $M\equiv M(m)$ is a $p\times p$ bandwidth matrix and $\mathbf{L}$ is a p-dimensional kernel function.

To conclude this first extension, and related to the proof of the asymptotic distribution, conditions (A.1) and (A.2) in Section 3 need to be generalised for a $p$-dimensional kernel function and a $p\times p$ bandwidth matrix, with the corresponding increasing complexity, specially in the selection procedures of those bandwidths.

\subsection{Non-Poisson point process}
It is well-known that there are practical situations in which Poisson models do not apply, because the interaction between points is not null, for example earthquakes and their aftershocks. In such scenarios, a point process can not be only characterised by its first-order intensity and the interaction, generally divided into attraction (producing clusters) and inhibition (originating regular patterns), needs to be determined.

Our idea is to extend our testing procedure to a more general class of point processes, is to conduct an extra step previous to the test in which we identify, and estimate, the type of interaction of our process. Once this is done, our test is well-defined and completely characterised through $H_0: \lambda(x)=\rho(Z(x)), \, x \in W$.

The definition of the statistic can be done in an analogous way, using the $L^2$-distance. The main barrier to overcome is its calibration; in our proposal we use the Poisson assumption in the bootstrap procedure performed to calibrate the distribution of the statistic and this will not be any more valid in these new scenarios. We propose to do this resampling procedure taking into account the previously modelled interaction: if for example we estimate it parametrically, we will have determined a certain type of process and then we can generate patterns (bootstrap resamples) from that model with the first-order intensity estimated non-parametrically using the proposal in \cite{Borrajo_Intensity}. Imagine we can stablish that we have for example a Neymann-Scott process, once we model the interaction, the only remaining part to model corresponds to the first-order intensity which will be enough to characterise the process and we can then formally perform our test over it.

Our last concern related to this extension on the nature of the process is that some of the theoretical results obtained under the Poisson assumption, may become much more harder to achieve or even intractable in such a general context. However we do not have those doubts on its practical applicability, using the ideas proposed in this section and applying Monte Carlo techniques as it has been done in \cite{DiazAvalosMateu2014}.

\section{Concluding remarks}
In this work we have used non-parametric techniques to define a test that allows to check the goodness-of-fit of an existing model in the field of point processes. Under such model the first-order intensity of a process can be expressed as a function of a spatial covariate. We propose a test for non-homogeneous Poisson point processes, where we have used the theoretical framework developed in \cite{Borrajo_Intensity} to detail the asymptotic normality of the test statistic as well as a bootstrap method used to accomplish its calibration. We have provided two real data examples to show the practicality of our proposal. We have also carried out a simulation study, based on these two real situations, to better analyse its performance. The results are in general really satisfactory: the simulations show competitive values in terms of level and power of the test in all the scenarios. Recall also that this procedure has no competitors, so no comparison with other methods is possible. To extend our proposal to a more general context we include a section with some possibilities: multidimensional covariates and non-Poisson processes. Gathering both ideas together allow us to define a really general test to determine whether the first-order intensity of any point process may be explained as a function of a certain $p$-dimensional covariate.

\section{Acknowledgements}
The authors acknowledge the support from the Spanish Ministry of Economy and Competitiveness, through grant number MTM2016-76969P, which includes support from the European Regional Development Fund (ERDF). Support from the IAP network StUDyS from Belgian Science Policy, is also acknowledged. M.I. Borrajo has been supported by FPU grant (FPU2013/00473) from the Spanish Ministry of Education. The authors also acknowledge the Canadian Wildland Fire Information System for their activity in recording and freely providing part of the real data used in this paper.

\bibliographystyle{plainnat}

\begin{thebibliography}{47}
	\providecommand{\natexlab}[1]{#1}
	\providecommand{\url}[1]{\texttt{#1}}
	\expandafter\ifx\csname urlstyle\endcsname\relax
	\providecommand{\doi}[1]{doi: #1}\else
	\providecommand{\doi}{doi: \begingroup \urlstyle{rm}\Url}\fi
	
	\bibitem[Baddeley and Van~Lieshout(1995)]{Baddeley1995}
	Adrian Baddeley and MNM Van~Lieshout.
	\newblock Area-interaction point processes.
	\newblock \emph{Annals of the Institute of Statistical Mathematics},
	47\penalty0 (4):\penalty0 601--619, 1995.
	\newblock \doi{10.1007/BF01856536}.
	
	\bibitem[Baddeley et~al.(2012)Baddeley, Chang, Song, and Turner]{Baddeley2012}
	Adrian Baddeley, Ya~Mei Chang, Yong Song, and Rolf. Turner.
	\newblock Nonparametric estimation of the dependence of a spatial point process
	on spatial covariates.
	\newblock \emph{Statistics and Its Interface}, 5:\penalty0 221--236, 2012.
	\newblock \doi{10.4310/SII.2012.v5.n2.a7}.
	
	\bibitem[Baddeley et~al.(2015)Baddeley, Rubak, and Turner]{Baddeley2015}
	Adrian Baddeley, Ege Rubak, and Rolf Turner.
	\newblock \emph{Spatial point patterns: methodology and applications with R}.
	\newblock CRC Press, 2015.
	
	\bibitem[Borrajo et~al.(2018)Borrajo, Gonz\'alez-Manteiga, and
	Mart\'inez-Miranda]{Borrajo_Intensity}
	Maria~I. Borrajo, W.~Gonz\'alez-Manteiga, and M.D. Mart\'inez-Miranda.
	\newblock Bootstrapping kernel intensity estimation for non-homogeneous point
	processes depending on spatial covariates.
	\newblock \emph{(Submitted) https://arxiv.org/pdf/1703.03213.pdf}, 2018.
	
	\bibitem[Borrajo et~al.(2017)Borrajo, Gonz{\'a}lez-Manteiga, and
	Mart{\'i}nez-Miranda]{Borrajo_LengthBiased}
	Mar{\'i}a~Isabel Borrajo, Wenceslao Gonz{\'a}lez-Manteiga, and
	Mar{\'i}a~Dolores Mart{\'i}nez-Miranda.
	\newblock Bandwidth selection for kernel density estimation for length-biased
	data.
	\newblock \emph{Journal of Nonparametric Statistics}, 29\penalty0 (3):\penalty0
	pp. 636--668, 2017.
	\newblock \doi{http://dx.doi.org/10.1080/10485252.2017.1339309}.
	
	\bibitem[Cao(1993)]{Cao1993}
	Ricardo Cao.
	\newblock Bootstrapping the mean integrated squared error.
	\newblock \emph{Journal of Multivariate Analysis}, 45\penalty0 (1):\penalty0
	137--160, 1993.
	\newblock \doi{10.1006/jmva.1993.1030}.
	
	\bibitem[Cowling et~al.(1996)Cowling, Hall, and Phillips]{Cowling1996}
	Ann Cowling, Peter Hall, and Michael~J Phillips.
	\newblock Bootstrap confidence regions for the intensity of a poisson point
	process.
	\newblock \emph{Journal of the American Statistical Association}, 91\penalty0
	(436):\penalty0 1516--1524, 1996.
	\newblock \doi{10.2307/2291577}.
	
	\bibitem[Cressie(2015)]{Cressiebook}
	Noel Cressie.
	\newblock \emph{Statistics for spatial data. Revised edition.}
	\newblock John Wiley \& Sons, 2015.
	\newblock \doi{10.1002/9781119115151}.
	
	\bibitem[Cucala(2006)]{CucalaThesis}
	Lionel Cucala.
	\newblock \emph{Espacements bidimensionnels et donn\'ees entach\'es d'erreurs
		dans l'analyse des procesus ponctuels spatiaux}.
	\newblock PhD thesis, Universit\'e des Sciences de Toulouse I, 2006.
	\newblock URL \url{https://tel.archives-ouvertes.fr/tel-00135890/document}.
	
	\bibitem[Daley and Vere-Jones(1988)]{DaleyVereJones1988}
	D.~J. Daley and D.~Vere-Jones.
	\newblock \emph{An introduction to the theory of point processes}.
	\newblock Springer Verlag, New York, 1988.
	
	\bibitem[Dasgupta and Raftery(1998)]{Dasgupta1998}
	Abhijit Dasgupta and Adrian~E Raftery.
	\newblock Detecting features in spatial point processes with clutter via
	model-based clustering.
	\newblock \emph{Journal of the American Statistical Association}, 93\penalty0
	(441):\penalty0 294--302, 1998.
	\newblock \doi{10.1080/01621459.1998.10474110}.
	
	\bibitem[Di~Iorio et~al.(2013)Di~Iorio, Anello, Bommarito, Cacciani, Denjean,
	De~Silvestri, Di~Biagio, di~Sarra, Ellul, Formenti,
	et~al.]{CanadaFires_SmokeMediterraneanSea}
	T~Di~Iorio, F~Anello, C~Bommarito, M~Cacciani, C~Denjean, L~De~Silvestri,
	C~Di~Biagio, A~di~Sarra, R~Ellul, P~Formenti, et~al.
	\newblock Long range transport of smoke particles from canadian forest fires to
	the mediterranean basin during june 2013.
	\newblock In \emph{AGU Fall Meeting Abstracts}, 2013.
	
	\bibitem[D{\'\i}az-Avalos et~al.(2014)D{\'\i}az-Avalos, Juan, and
	Mateu]{DiazAvalosMateu2014}
	Carlos D{\'\i}az-Avalos, P~Juan, and Jorge Mateu.
	\newblock Significance tests for covariate-dependent trends in inhomogeneous
	spatio-temporal point processes.
	\newblock \emph{Stochastic environmental research and risk assessment},
	28\penalty0 (3):\penalty0 593--609, 2014.
	\newblock \doi{10.1007/s00477-013-0775-1}.
	
	\bibitem[Diggle and Marron(1988)]{DiggleMarron1988}
	Peter Diggle and James~Stephen Marron.
	\newblock Equivalence of smoothing parameter selectors in density and intensity
	estimation.
	\newblock \emph{Journal of the American Statistical Association}, 83:\penalty0
	pp. 793--800, 1988.
	\newblock \doi{10.2307/2289308}.
	
	\bibitem[Diggle(1985)]{Diggle1985}
	Peter~J. Diggle.
	\newblock A kernel method for smoothing point process data.
	\newblock \emph{Journal of the Royal Statistical Society. Series C (Applied
		Statistics)}, 34\penalty0 (2):\penalty0 138--147, 1985.
	\newblock \doi{10.2307/2347366}.
	
	\bibitem[Diggle(2013)]{Diggle2013}
	Peter~J Diggle.
	\newblock \emph{Statistical analysis of spatial and spatio-temporal point
		patterns}.
	\newblock CRC Press, 2013.
	\newblock \doi{10.1002/bimj.201400024}.
	
	\bibitem[Duong(2013)]{Duong2013}
	Tarn Duong.
	\newblock Local significant differences from nonparametric two-sample tests.
	\newblock \emph{Journal of Nonparametric Statistics}, 25\penalty0 (3):\penalty0
	635--645, 2013.
	\newblock \doi{10.1080/10485252.2013.810217}.
	
	\bibitem[Eubank and Hart(1992)]{Hart1992}
	RL~Eubank and Jeffrey~D Hart.
	\newblock Testing goodness-of-fit in regression via order selection criteria.
	\newblock \emph{The annals of Statistics}, 20\penalty0 (3):\penalty0
	1412--1425, 1992.
	
	\bibitem[Flannigan and Harrington(1988)]{CanadaFires_Meteorological}
	MD~Flannigan and J\_B Harrington.
	\newblock A study of the relation of meteorological variables to monthly
	provincial area burned by wildfire in canada (1953--80).
	\newblock \emph{Journal of Applied Meteorology}, 27\penalty0 (4):\penalty0
	441--452, 1988.
	\newblock \doi{10.1175/1520-0450(1988)027<0441:ASOTRO>2.0.CO;2}.
	
	\bibitem[Foxall and Baddeley(2002)]{foxall2002nonparametric}
	Rob Foxall and Adrian Baddeley.
	\newblock Nonparametric measures of association between a spatial point process
	and a random set, with geological applications.
	\newblock \emph{Journal of the Royal Statistical Society: Series C (Applied
		Statistics)}, 51\penalty0 (2):\penalty0 165--182, 2002.
	
	\bibitem[Fuentes-Santos et~al.(2015)Fuentes-Santos, Gonz{\'a}lez-Manteiga, and
	Mateu]{Isa2015}
	Isabel Fuentes-Santos, Wenceslao Gonz{\'a}lez-Manteiga, and Jorge Mateu.
	\newblock Consistent smooth bootstrap kernel intensity estimation for
	inhomogeneous spatial poisson point processes.
	\newblock \emph{Scandinavian Journal of Statistics}, 43\penalty0 (2):\penalty0
	416--435, 2015.
	\newblock \doi{10.1111/sjos.12183}.
	
	\bibitem[Fuentes-Santos et~al.(2017)Fuentes-Santos, Gonz{\'a}lez-Manteiga, and
	Mateu]{IsaSpatStat}
	Isabel Fuentes-Santos, Wenceslao Gonz{\'a}lez-Manteiga, and Jorge Mateu.
	\newblock A nonparametric test for the comparison of first-order structures of
	spatial point processes.
	\newblock \emph{Spatial Statistics}, 22:\penalty0 240--260, 2017.
	\newblock \doi{10.1016/j.spasta.2017.02.007}.
	
	\bibitem[Garc\'ia-Portugu\'es et~al.(2016)Garc\'ia-Portugu\'es, Van~Keilegom,
	Crujeiras, and Gonz\'alez-Manteiga]{EduIngrid}
	E.~Garc\'ia-Portugu\'es, I.~Van~Keilegom, R.M. Crujeiras, and
	W.~Gonz\'alez-Manteiga.
	\newblock Testing parametric models in linear-directional regression.
	\newblock \emph{Scand. J. Statist.}, 43\penalty0 (4):\penalty0 1178--1191,
	2016.
	
	\bibitem[Guan(2008)]{Guan2008}
	Yongtao Guan.
	\newblock On consistent nonparametric intensity estimation for inhomogeneous
	spatial point processes.
	\newblock \emph{Journal of the American Statistical Association}, 103\penalty0
	(483):\penalty0 1238--1247, 2008.
	\newblock \doi{10.1198/016214508000000526}.
	
	\bibitem[Guan and Loh(2007)]{GuanLoh2007}
	Yongtao Guan and Ji~Meng Loh.
	\newblock A thinned block bootstrap variance estimation procedure for
	inhomogeneous spatial point patterns.
	\newblock \emph{Journal of the American Statistical Association}, 102\penalty0
	(480):\penalty0 1377--1386, 2007.
	
	\bibitem[Guan and Shen(2010)]{GuanShen2010}
	Yongtao Guan and Ye~Shen.
	\newblock A weighted estimating equation approach for inhomogeneous spatial
	point processes.
	\newblock \emph{Biometrika}, 97\penalty0 (4):\penalty0 867--880, 2010.
	
	\bibitem[Hall(1984)]{Hall1984}
	Peter Hall.
	\newblock Central limit theorem for integrated square error of multivariate
	nonparametric density estimators.
	\newblock \emph{Journal of multivariate analysis}, 14\penalty0 (1):\penalty0
	1--16, 1984.
	\newblock \doi{10.1016/0047-259X(84)90044-7}.
	
	\bibitem[Hyndman and Yao(2002)]{Hyndman2002}
	Rob~J Hyndman and Qiwei Yao.
	\newblock Nonparametric estimation and symmetry tests for conditional density
	functions.
	\newblock \emph{Journal of nonparametric statistics}, 14\penalty0 (3):\penalty0
	259--278, 2002.
	
	\bibitem[Illian et~al.(2009)Illian, M{\o}ller, and Waagepetersen]{Illian2009}
	Janine~B Illian, Jesper M{\o}ller, and Rasmus~P Waagepetersen.
	\newblock Hierarchical spatial point process analysis for a plant community
	with high biodiversity.
	\newblock \emph{Environmental and Ecological Statistics}, 16\penalty0
	(3):\penalty0 389--405, 2009.
	\newblock \doi{10.1007/s10651-007-0070-8}.
	
	\bibitem[Jones(1991)]{Jones1991}
	Michael~C Jones.
	\newblock Kernel density estimation for length biased data.
	\newblock \emph{Biometrika}, 78\penalty0 (3):\penalty0 pp. 511--519, 1991.
	\newblock \doi{10.1093/biomet/78.3.511}.
	
	\bibitem[Law et~al.(2009)Law, Illian, Burslem, Gratzer, Gunatilleke, and
	Gunatilleke]{Law2009}
	Richard Law, Janine~B Illian, David~FRP Burslem, Georg Gratzer, CVS
	Gunatilleke, and IAUN Gunatilleke.
	\newblock Ecological information from spatial patterns of plants: insights from
	point process theory.
	\newblock \emph{Journal of Ecology}, 97\penalty0 (4):\penalty0 616--628, 2009.
	\newblock \doi{10.1111/j.1365-2745.2009.01510.x}.
	
	\bibitem[Lawson(2013)]{Lawson2013}
	Andrew~B Lawson.
	\newblock \emph{Statistical methods in spatial epidemiology}.
	\newblock John Wiley \& Sons, 2013.
	\newblock \doi{10.1002/9780470035771.biblio}.
	
	\bibitem[Mammen et~al.(1992)Mammen, Marron, and Fisher]{MammenMarronFisher1992}
	Enno Mammen, James~S Marron, and Nick~I Fisher.
	\newblock Some asymptotics for multimodality tests based on kernel density
	estimates.
	\newblock \emph{Probability Theory and Related Fields}, 91\penalty0
	(1):\penalty0 115--132, 1992.
	
	\bibitem[M{\o}ller and Waagepetersen(2003)]{Moller2003}
	Jesper M{\o}ller and Rasmus~Plenge Waagepetersen.
	\newblock \emph{Statistical inference and simulation for spatial point
		processes}.
	\newblock CRC Press, 2003.
	\newblock \doi{10.1002/sim.1896}.
	
	\bibitem[Ogata and Zhuang(2006)]{OgataZhuang2006}
	Yosihiko Ogata and Jiancang Zhuang.
	\newblock Space--time etas models and an improved extension.
	\newblock \emph{Tectonophysics}, 413\penalty0 (1):\penalty0 13--23, 2006.
	\newblock \doi{10.1016/j.tecto.2005.10.016}.
	
	\bibitem[Reitzner and Schulte(2013)]{Reitzner2013}
	Matthias Reitzner and Matthias Schulte.
	\newblock Central limit theorems for $ u $-statistics of poisson point
	processes.
	\newblock \emph{The Annals of Probability}, 41\penalty0 (6):\penalty0
	3879--3909, 2013.
	\newblock \doi{10.1214/12-AOP817}.
	
	\bibitem[Rogers et~al.(2013)Rogers, Randerson, and
	Bonan]{CanadaFires_HighLatitudeCooling}
	BM~Rogers, JT~Randerson, and GB~Bonan.
	\newblock High-latitude cooling associated with landscape changes from north
	american boreal forest fires.
	\newblock \emph{Biogeosciences}, 10\penalty0 (2):\penalty0 699--718, 2013.
	\newblock \doi{10.5194/bg-10-699-2013}.
	
	\bibitem[Schoenberg(2005)]{Schoenberg2005}
	Frederic~Paik Schoenberg.
	\newblock Consistent parametric estimation of the intensity of a
	spatial--temporal point process.
	\newblock \emph{Journal of Statistical Planning and Inference}, 128\penalty0
	(1):\penalty0 79--93, 2005.
	
	\bibitem[Schoenberg(2011)]{Schoenberg2011}
	Frederic~Paik Schoenberg.
	\newblock Multidimensional residual analysis of point process models for
	earthquake occurrences.
	\newblock \emph{Journal of the American Statistical Association}, 98:\penalty0
	789--795, 2011.
	\newblock \doi{10.1198/016214503000000710}.
	
	\bibitem[Silverman(1981)]{Silverman1981}
	Bernard~W Silverman.
	\newblock Using kernel density estimates to investigate multimodality.
	\newblock \emph{Journal of the Royal Statistical Society. Series B
		(Methodological)}, 4\penalty0 (1):\penalty0 97--99, 1981.
	
	\bibitem[Stoyan and Penttinen(2000)]{Stoyan2000}
	Dietrich Stoyan and Antti Penttinen.
	\newblock Recent applications of point process methods in forestry statistics.
	\newblock \emph{Statistical Science}, 15\penalty0 (1):\penalty0 61--78, 2000.
	\newblock \doi{10.1214/ss/1009212674}.
	
	\bibitem[Thurman et~al.(2015)Thurman, Fu, Guan, and Zhu]{ThurmanGuan2015}
	Andrew~L Thurman, Rao Fu, Yongtao Guan, and Jun Zhu.
	\newblock Regularized estimating equations for model selection of clustered
	spatial point processes.
	\newblock \emph{Statistica Sinica}, 25\penalty0 (1):\penalty0 173--188, 2015.
	
	\bibitem[Van~Lieshout(2000)]{Vanlis2000}
	MNM Van~Lieshout.
	\newblock \emph{Markov point processes and their applications}.
	\newblock World Scientific, 2000.
	\newblock \doi{10.1142/9781860949760}.
	
	\bibitem[Waagepetersen(2007)]{Waagepetersen2007}
	Rasmus~Plenge Waagepetersen.
	\newblock An estimating function approach to inference for inhomogeneous
	neyman--scott processes.
	\newblock \emph{Biometrics}, 63\penalty0 (1):\penalty0 252--258, 2007.
	\newblock \doi{10.1111/j.1541-0420.2006.00667.x}.
	
	\bibitem[Walter et~al.(2014)Walter, Freitas, Kraut, Rieger, Vogel, and
	Vogel]{CanadaFires_Clouds}
	C~Walter, SR~Freitas, I~Kraut, D~Rieger, H~Vogel, and B~Vogel.
	\newblock Influence of 2010 canadian forest fires on cloud formation on the
	regional scale.
	\newblock In \emph{AGU Fall Meeting Abstracts}, 2014.
	
	\bibitem[Watkins and Hickman(1990)]{MurchisonSurvey}
	Keith~P Watkins and Arthur~Hugh Hickman.
	\newblock \emph{Geological evolution and mineralization of the Murchison
		Province, Western Australia}, volume~1.
	\newblock Department of Mines, Western Australia, 1990.
	
	\bibitem[Yue and Loh(2015)]{YueLoh2015}
	Yu~Ryan Yue and Ji~Meng Loh.
	\newblock Variable selection for inhomogeneous spatial point process models.
	\newblock \emph{Canadian Journal of Statistics}, 43\penalty0 (2):\penalty0
	288--305, 2015.
	
\end{thebibliography}

\newpage
\appendix
\renewcommand{\thesection}{Appendix \Alph{section} -}
\section{Proof of Theorem \ref{th:normality}}
Along this proof we will obtain the mean and variance of the statistic $T$ as well as assuring its asymptotic normality. To deal with this statistic we will rewrite it in an more suitable way:
\begin{align}\label{eq:Texpresion}
	T&=\int_W{\left(\hat{\lambda}_{0,H}(x)-\hat{\rho}_{0,b}(Z(x))\right)^2dx}=\int_W{\left(\hat{\lambda}_{0,H}(x)-\lambda_0(x)\right)^2dx}+ \nonumber \\ &+\int_W{\left(\lambda_0(x)-\hat{\rho}_{0,b}(Z(x))\right)^2dx} + \int_W{\left(\hat{\lambda}_{0,H}(x)-\lambda_0(x)\right)\left(\lambda_0(x)-\hat{\rho}_{0,b}(Z(x))\right)dx.}
	\end{align}

\noindent \underline{\textit{Mean and variance of T}}\\

Taking \eqref{eq:Texpresion} into account, we immediately obtain that  
\begin{align}
	E\left[T\right]&=E\left[\int_W{\left(\hat{\lambda}_{0,H}(x)-\lambda_0(x)\right)^2dx}\right] + E\left[\int_W{\left(\lambda_0(x)-\hat{\rho}_{0,b}(Z(x))\right)^2dx}\right] + \nonumber \\
	&+E\left[\int_W{\left(\hat{\lambda}_{0,H}(x)-\lambda_0(x)\right)\left(\lambda_0(x)-\hat{\rho}_{0,b}(Z(x))\right)dx}\right]
	\end{align}

and

	\begin{align}
	Var\left[T\right]&=Var\left[\int_W{\left(\hat{\lambda}_{0,H}(x)-\lambda_0(x)\right)^2dx}\right] + Var\left[\int_W{\left(\lambda_0(x)-\hat{\rho}_{0,b}(Z(x))\right)^2dx}\right] + \nonumber \\
	&+Var\left[\int_W{\left(\hat{\lambda}_{0,H}(x)-\lambda_0(x)\right)\left(\lambda_0(x)-\hat{\rho}_{0,b}(Z(x))\right)dx}\right] +\nonumber \\
	&+ 2Cov\left(\int_W{\left(\hat{\lambda}_{0,H}(x)-\lambda_0(x)\right)^2dx},\int_W{\left(\lambda_0(x)-\hat{\rho}_{0,b}(Z(x))\right)^2dx}\right)+\nonumber \\ &+2Cov\left(\int_W{\left(\hat{\lambda}_{0,H}(x)-\lambda_0(x)\right)^2dx},\int_W{\left(\hat{\lambda}_{0,H}(x)-\lambda_0(x)\right)\left(\lambda_0(x)-\hat{\rho}_{0,b}(Z(x))\right)dx}\right) + \nonumber \\
	&+ 2Cov(\int_W{\left(\lambda_0(x)-\hat{\rho}_{0,b}(Z(x))\right)^2dx},\int_W{\left(\hat{\lambda}_{0,H}(x)-\lambda_0(x)\right)\left(\lambda_0(x)-\hat{\rho}_{0,b}(Z(x))\right)dx}).
	\end{align}

So, we first of all compute the mean and the variance of each of the addends in \eqref{eq:Texpresion}, and finally we will deal with the covariances between the different terms. For all of them, the mathematical tools we apply consist of first obtain the explicit expressions for the squares, then swap the mean operator and the integrals, and then compute several means of product terms of the involved estimates. In this last step we use properties of conditional mean as well as some Taylor's expansions.\\

\noindent\textit{First addend}\\
\begin{align}\label{eq:mean1}
	E\left[\int_W{\left(\hat{\lambda}_{0,H}(x)-\lambda_0(x)\right)^2dx}\right]&=\int E\left[\hat{\lambda}_{0,H}(x)\right]dx -2\int{\lambda_0(x)\hat{\lambda}_{0,H}(x)dx} + R(\lambda_0)\nonumber \\
	&= A(m)|H|^{-1/2}R(K) + \frac{1}{4}\mu_2^2(K)\int{tr^2(HD^2\lambda_0(x))dx}\nonumber \\
	& + o\left(A(m)|H|^{-1/2}\right) + o\left(tr(H)\right)
	\end{align}
and
\begin{align}\label{eq:var1}
	Var\left[\int_W{\left(\hat{\lambda}_{0,H}(x)-\lambda_0(x)\right)^2dx}\right]&= \int\int\left(E\left[\hat{\lambda}^2_{0,H}(x)\hat{\lambda}^2_{0,H}(y)\right] + 2\lambda^2_0(y)E\left[\hat{\lambda}^2_{0,H}(x)\right]\right. \nonumber \\ &\left.-4\lambda_0(y)E\left[\hat{\lambda}^2_{0,H}(x)\hat{\lambda}_{0,H}(y)\right] - 4\lambda^2_0(x)\lambda_0(y)E\left[\hat{\lambda}_{0,H}(y)\right]\right. \nonumber \\
	&\left. + 4\lambda_0(x)\lambda_0(y)E\left[\hat{\lambda}_{0,H}(x)\hat{\lambda}_{0,H}(y)\right] + \lambda^2_0(x)\lambda^2_0(y)\right)dxdy\nonumber \\
	& = R(K)o(A(m)|H|^{-1/2}) - 2R(K)R(\lambda_0)o(A(m)|H|^{-1/2}) \nonumber \\
	&- 6R(\lambda_0)o(tr(H))
	\end{align}
where we have used the following equations:
\begin{align*}
	E\left[K_H(x-X_1)\right]=\int{K_H(x-u)\lambda_0(u)du}=\lambda_0(x) + \frac{1}{2}\mu_2(K)\trdsegx + o(tr(H)),
	\end{align*}
	
	\begin{align*}
	E\left[K^2_H(x-X_1)\right]=\int{K^2_H(x-u)\lambda_0(u)du}=|H|^{-1/2}\lambda_0(x)R(K) + o(|H|^{-1/2}),
	\end{align*}
	
	\begin{align*}
	E\left[K_H(x-X_1)K_H(y-X-1)\right]&=\int{K_H(x-u)K_H(y-u)\lambda_0(u)du}=\nonumber \\
	&=|H|^{-1/2}\lambda_0(x)\KK + o(|H|^{-1/2}),
	\end{align*}
	
	\begin{align*}
	E\left[K^2_H(x-X_1)K_H(y-X_1)\right]&=\int{K^2_H(x-u)K_H(y-u)\lambda_0(u)du}=\nonumber \\
	&=|H|^{-1}\lambda_0(x)\KKc + o(|H|^{-1}),
	\end{align*}
	
	\begin{align*}
	E\left[K^2_H(x-X_1)K^2_H(y-X_1)\right]&=\int{K^2_H(x-u)K^2_H(y-u)\lambda_0(u)du}=\nonumber \\
	&=|H|^{-3/2}\lambda_0(x)\KKcc + o(|H|^{-3/2}),
	\end{align*}
	
	\begin{align*}
	& E\left[\hat{\lambda}_{0,H}(x)\right]=(1-e^{-m})E\left[K_H(x-X_1)\right]=\lambda_0(x) + \frac{1}{2}\mu_2(K)\trdsegx + o(tr(H)), \qquad
	\end{align*}
	
	\begin{align*}
	E\left[\hat{\lambda}^2_{0,H}(x)\right]&=A(m)E\left[K^2_H(x-X_1)\right]+(1-e^{-m}-A(m))E^2\left[K_H(x-X_1)\right]\nonumber \\
	&= A(m)|H|^{-1/2}\lambda_0(x)R(K) + \lambda_0^2(x)+\mu_2(K)\lambda_0(x)\trdsegx\nonumber \\
	&+\frac{1}{4}\mu_2^2(K)\trdsegxc+A(m)\lambda_0^2(x)+A(m)\mu_2(K)\lambda_0(x)\trdsegx \nonumber \\
	& + \frac{1}{4}A(m)\mu_2^2(K)\trdsegxc+ 2\lambda_0(x)o(tr(H))+R(K)o(A(m)|H|^{-1/2})\\
	&+\mu_2(K)\trdsegx o(tr(H))+o(tr^2(H))+ o(A(m))
	\end{align*}
	
	\begin{align*}
	E\left[\hat{\lambda}_{0,H}(x)\hat{\lambda}_{0,H}(y)\right]&=A(m)E\left[K_H(x-X_1)K_H(y-X_1)\right]\nonumber \\
	&+(1-e^{-m}-A(m))E\left[K_H(x-X_1)\right]E\left[K_H(y-X_1)\right]\\
	&= A(m)|H|^{-1/2}\lambda_0(x)\KK +\nonumber \\
	&+ \KK o(A(m)|H|^{-1/2}) + \lambda_0(x)\lambda_0(y)\\
	& + \frac{1}{2}\lambda_0(x)\mu_2(K)\trdsegy\lambda_0(x)o(tr(H))\\
	&+\frac{1}{2}\lambda_0(y)\mu_2(K)\trdsegx+\frac{1}{4}\mu_2^2(K)\trdsegx\trdsegy\nonumber \\ 
	&+\frac{1}{2}\mu_2(K)\trdsegx o(tr(H))+\lambda_0(y)o(tr(H))\\
	&+\frac{1}{2}\mu_2(K)\trdsegy o(tr(H)) + A(m)\lambda_0(x)\lambda_0(y)\\
	&+\frac{1}{2}A(m)\lambda_0(x)\mu_2(K)\trdsegy + A(m)\lambda_0(x)o(tr(H))+\nonumber \\
	&+\frac{1}{2}\lambda_0(y)\mu_2(K)\trdsegx \\
	&+ \frac{1}{4}A(m)\mu_2^2(K)\trdsegx\trdsegy+ o(A(m)tr(H)),
	\end{align*}
	
	\begin{align*}
	E\left[\hat{\lambda}^2_{0,H}(x)\hat{\lambda}_{0,H}(y)\right]&=B(m)E\left[K^2_H(x-X_1)K_H(y-X_1)\right]\nonumber \\ &+A(m)E\left[K^2_H(x-X_1)\right]E\left[K_H(y-X_1)\right]\\
	&+2A(m)E\left[K_H(x-X_1)K_H(y-X_1)\right]E\left[K_H(x-X_1)\right]\nonumber \\
	&+(1-e^{-m})E^2\left[K_H(x-X_1)\right]E\left[K_H(y-X_1)\right]\\
	&= A(m)|H|^{-1/2}\lambda_0(x)\lambda_0(y)R(K)  \nonumber \\
	&+\frac{1}{2}A(m)|H|^{-1/2}\lambda_0(x)\mu_2(K)R(K)\trdsegx\\
	& + \lambda_0(x)R(K)o(A(m)|H|^{-1/2}tr(H)) + \lambda_0(y)R(K)o(A(m)|H|^{-1/2})\\
	&+\frac{1}{2}R(K)\mu_2(K)\trdsegy o(A(m)|H|^{-1/2})  \nonumber \\
	&+ R(K)o(A(m)|H|^{-1/2}tr(H)) +2A(m)|H|^{-1/2}\lambda^2_0(x)\KK \nonumber \\
	&+ A(m)|H|^{-1/2}\mu_2(K)\KK\lambda_0(x)\trdsegx \nonumber \\
	&+ 2\KK o(A(m)|H|^{-1/2}tr(H))\\
	& +2\lambda_0(x)\KK o(A(m)|H|^{-1/2}) \nonumber \\
	&+ \mu_2(K)\KK \trdsegx o(A(m)|H|^{-1/2}) \\
	&+ o(A(m)|H|^{-1/2}tr(H)) +\lambda^2_0(x)\lambda_0(y) + \frac{1}{2}\mu_2(K)\lambda^2_0(x)\trdsegy\\
	& + \lambda^2_0(x)o(tr(H))+\mu_2(K)\lambda_0(x)\lambda_0(y)\trdsegx+\nonumber \\ &+\frac{1}{2}\mu_2^2(K)\lambda_0(x)\trdsegx\trdsegy\\
	&+\mu_2(K)\lambda_0(x)\trdsegx o(tr(H))+2\lambda_0(x)\lambda_0(y)p(tr(H))\\
	& + \mu_2(K)\lambda_0(x)\trdsegy o(tr(H)) +2\lambda_0(x)o(tr^2(H))+\lambda_0(y)o(tr^2(H))\\
	&+\frac{1}{2}\mu_2(K)\trdsegy o(tr^2(H))+o(tr^3(H)),
	\end{align*}
	
	\begin{align*}
	E\left[\hat{\lambda}^2_{0,H}(x)\hat{\lambda}^2_{0,H}(y)\right]&=C(m)E\left[K^2_H(x-X_1)K^2_H(y-X_1)\right] \nonumber \\ 
	&+ 2B(m)E\left[K^2_H(x-X_1)K_H(y-X_1)\right]E\left[K_H(y-X_1)\right]\nonumber \\ 
	&+2B(m)E\left[K_H(x-X_1)K^2_H(y-X_1)\right]E\left[K_H(x-X_1)\right]\nonumber \\ 
	&+B(m)E\left[K^2_H(x-X_1)\right]E\left[K^2_H(y-X_1)\right]\\
	&+2B(m)E^2\left[K_H(x-X_1)K_H(y-X_1)\right] \nonumber \\ 
	&+ A(m)E\left[K^2_H(x-X_1)\right]E^2\left[K_H(y-X_1)\right] \nonumber \\ 
	&+ 4A(m)E\left[K_H(x-X_1)K_H(y-X_1)\right]E\left[K_H(x-X_1)\right]E\left[K_H(y-X_1)\right] \nonumber \\ 
	&+ A(m)E\left[K^2_H(y-X_1)\right]E^2\left[K_H(x-X_1)\right]\\
	& + (1-e^{-m})E^2\left[K_H(x-X_1)\right]E^2\left[K_H(y-X_1)\right]=\nonumber \\ 
	&= A(m)|H|^{-1/2}\lambda_0(x)\lambda^2_0(y)R(K)\\
	& + A(m)|H|^{-1/2}R(K)\mu_2(K)\lambda_0(x)\lambda_0(y)\trdsegy\nonumber \\ &+R(K)\lambda^2_0(y)o(A(m)|H|^{-1/2})+R(K)\mu_2(K)\lambda_0(y)\trdsegy o(A(m)|H|^{-1/2}) \nonumber \\ 
	&+ 4A(m)|H|^{-1/2}\lambda^2_0(x)\lambda_0(y)\KK \nonumber \\ 
	&+ 2A(m)|H|^{-1/2}\lambda^2_0(x)\mu_2(K)\KK\trdsegy\nonumber \\ 
	&+4A(m)\lambda^2_0(x)\KK o(|H|^{-1/2}tr(H))\nonumber \\ 
	&+2A(m)|H|^{-1/2}\lambda_0(x)\lambda_0(y)\mu_2(K)\KK\trdsegx+\nonumber \\ 
	&+4\lambda_0(x)\lambda_0(y)\KK o(A(m)|H|^{-1/2})\\
	& + o(A(m)|H|^{-1/2}tr(H))+  A(m)|H|^{-1/2}\lambda^2_0(x)\lambda_0(y)R(K)\\
	&+A(m)|H|^{-1/2}R(K)\mu_2(K)\lambda_0(x)\lambda_0(y)\trdsegx\nonumber \\ 
	&+ R(K)\lambda^2_0(x)o(A(m)|H|^{-1/2})\\
	&+R(K)\mu_2(K)\lambda_0(x)\trdsegx o(A(m)|H|^{-1/2})\\
	& + \lambda^2_0(x)\lambda^2_0(y) + \lambda^2_0(x)\lambda_0(y)\mu_2(K)\trdsegy 2\lambda^2_0(x)\lambda_0(y)o(tr(H))\\
	&+\frac{1}{4}\lambda^2_0(x)\mu_2^2(K)\trdsegyc +\lambda^2_0(x)\mu_2(K)\trdsegy o(tr(H))\\
	& + \lambda^2_0(y)\lambda_0(x)\mu_2(K)\trdsegx \nonumber \\ 
	&+ \mu_2^2(K)\lambda_0(x)\lambda_0(y)\trdsegx\trdsegy\\
	& + 2\mu_2(K)\lambda_0(x)\lambda_0(y)\trdsegx o(tr(H)) \nonumber \\ 
	&+ 2\lambda_0(x)\lambda^2_0(y)o(tr(H))+2\mu_2(K)\lambda_0(x)\lambda_0(y)\trdsegy o(tr(H)) \nonumber \\ 
	&+ \frac{1}{4}\mu_2^2(K)\lambda^2_0(y)\trdsegxc + \lambda^2_0(y)\mu_2(K)\trdsegx o(tr(H))\\
	& +o(A(m))+o(tr^2(H)),
	\end{align*}
with $B(m)=E\left[\frac{1}{N^2}1_{\{N\neq0\}}\right]$ and $C(m)=E\left[\frac{1}{N^3}1_{\{N\neq0\}}\right]$.\\

\noindent\textit{Second addend}\\
We have also computed the mean and variance of the second addend using the same tools as in the first one, and also the relationship established in Theorem A.1 and Theorem A.2 in \cite{Borrajo_Intensity}. We finally obtain that both, mean and variance, are negligible in comparison with the terms obtained for the first addend, in particular we have found that 
\[E\left[\int_W{\left(\lambda_0(x)-\hat{\rho}_{0,b}(Z(x))\right)^2dx}\right]=o(A(m))\]
and
\[Var\left[\int_W{\left(\lambda_0(x)-\hat{\rho}_{0,b}(Z(x))\right)^2dx}\right]=0(A(m)),\]
which are both smaller than $o(A(m)|H|^{-1/2}+tr(H))$ corresponding to the first addend.\\

\noindent\textit{Third addend}\\
\begin{align*}
	E\left[\int_W{\left(\hat{\lambda}_{0,H}(x)-\lambda_0(x)\right)\left(\lambda_0(x)-\hat{\rho}_{0,b}(Z(x))\right)dx}\right]&=\frac{1}{2}\mu_2(K)\int{\lambda_0(x)\trdsegx dx}\\
	&+o(tr(H))
	\end{align*}

\noindent and 

\begin{align*}
	& Var\left[\int_W{\left(\hat{\lambda}_{0,H}(x)-\lambda_0(x)\right)\left(\lambda_0(x)-\hat{\rho}_{0,b}(Z(x))\right)dx}\right]=\nonumber  \\
	&=A(m)|H|^{-1/2}\int\int{\lambda^2_0(x)\lambda_0(y)\KK dx dy} +\nonumber \\
	&+ \int\int{\lambda_0(x)\lambda_0(y)\KK o(A(m)|H|^{-1/2})dx dy} + o(A(m)),
	\end{align*}

\noindent where we have used that $E\left[\hat{\rho}_{0,b}(x)\right]$, $E\left[\hat{\rho}_{0,b}(x)\hat{\rho}_{0,b}(y)\right]$, $E\left[\hat{\lambda}_{0,H}(x)\hat{\rho}_{0,b}(x)\right]$, \linebreak $E\left[\hat{\lambda}_{0,H}(x)\hat{\rho}_{0,b}(x)\hat{\rho}_{0,b}(y)\right]$, $E\left[\hat{\lambda}_{0,H}(x)\hat{\lambda}_{0,H}(y)\hat{\rho}_{0,b}(x)\right]$ and \linebreak  $E\left[\hat{\lambda}_{0,H}(x)\hat{\lambda}_{0,H}(y)\hat{\rho}_{0,b}(x)\hat{\rho}_{0,b}(y)\right]$ are smaller than the main term in the first addend's variance. And then the only terms that still contribute here to the global variance are: $E\left[\hat{\lambda}_{0,H}(x)\right]$ and $E\left[\hat{\lambda}_{0,H}(x)\hat{\lambda}_{0,H}(y)\right]$, which have already been calculated in a previous step of this proof.\\

\noindent\textit{Covariances}\\

\begin{align}
	& Cov\left[\int_W{\left(\hat{\lambda}_{0,H}(x)-\lambda_0(x)\right)^2dx},\int_W{\left(\lambda_0(x)-\hat{\rho}_{0,b}(Z(x))\right)^2dx}\right]=\nonumber \\
	& =A(m)|H|^{-1/2}R(\lambda_0)R(K)+ R(\lambda_0)R(K)o(A(m)|H|^{-1/2}) + o(A(m))
	\end{align}

\noindent and 

\begin{align}
	Cov\left[\int_W{\left(\hat{\lambda}_{0,H}(x)-\lambda_0(x)\right)^2dx},\int_W{\left(\hat{\lambda}_{0,H}(x)-\lambda_0(x)\right)\left(\lambda_0(x)-\hat{\rho}_{0,b}(Z(x))\right)dx}\right]
	\end{align}
\noindent and
\begin{align}
	Cov\left[\int_W{\left(\lambda_0(x)-\hat{\rho}_{0,b}(Z(x))\right)^2dx},\int_W{\left(\hat{\lambda}_{0,H}(x)-\lambda_0(x)\right)\left(\lambda_0(x)-\hat{\rho}_{0,b}(Z(x))\right)dx}\right]
	\end{align}
are smaller than the main term of the first addend's variance.\\

\noindent \underline{\textit{Asymptotic normality}}\\
Our test statistic can be expanded and written as:
\begin{align}\label{eq:Tustat}
	T&=\int_W{\left(\hat{\lambda}_{0,H}(x)-\hat{\rho}_{0,b}(Z(x))\right)^2dx}=\frac{1}{N^2}\sum_{i=1}^N\int{K_H(x-X_i)dx}+\nonumber \\
	&+\frac{1}{N^2}\sum_{i=1}^N\sum_{j\neq i}\int{K_H(x-X_i)K_H(x-X_j)dx}++\nonumber \\
	&+\frac{1}{N^2}\sum_{i=1}^N\int{\frac{1}{g^\star(Z(X_i))}L_b(Z(x)-Z(X_i))dx}+\nonumber \\
	&+\frac{1}{N^2}\sum_{i=1}^N\sum_{j\neq i}\int{\frac{1}{g^\star(Z(X_i))}\frac{1}{g^\star(Z(X_j))}L_b(x-X_i)L_b(x-X_j)dx}-\nonumber \\
	&-\frac{2}{N^2}\sum_{i=1}^N\int{\frac{1}{g^\star(Z(X_i))}K_H(x-X_i)L_b(Z(x)-Z(X_i))dx}-\nonumber \\
	&-\frac{2}{N^2}\sum_{i=1}^N\sum_{j\neq i}\int{\frac{1}{g^\star(Z(X_j))}K_H(x-X_i)L_b(Z(x)-Z(X_j))dx}
	\end{align}
	where each of the addends is a U-statistic on a Poisson point process, remarking that the sums does not allow duplicated points in the same expression. Moreover, every of the addends is absolutely convergent in the sense defined by \cite{Reitzner2013}, hence following their Theorem 4.7 we can assure the normality of each term. Then the normality of our statistic with the mean and variance detailed in the main body of Theorem \ref{th:normality} is proved.

\end{document}